\documentclass[aps,prd,floatfix,notitlepage,superscriptaddress,preprintnumbers,nofootinbib]{revtex4-2}
\pdfoutput=1
\usepackage{setspace}
\usepackage{amsmath,amssymb,amsfonts}
\usepackage{color}
\usepackage{subfigure} 
\usepackage{natbib} 
\definecolor{darkblue}{rgb}{0,0,0.5}
\usepackage[colorlinks,linkcolor=blue,citecolor=blue,urlcolor=blue]{hyperref}

\allowdisplaybreaks

\newcommand\beq{\begin{equation}}
\newcommand\eeq{\end{equation}}
\newcommand\Tr{\text{Tr}\,}

\newcommand\hc{\text{H.c.}}

\usepackage{rotating}
\usepackage{tabularx}
\usepackage{multirow}
\usepackage{xurl}
\usepackage{physics}
\usepackage{xparse}
\usepackage{enumitem}
\usepackage{resizegather}
\usepackage{float}

\usepackage[many]{tcolorbox}

\newcommand{\eq}[1]{Eq.~\eqref{#1}}
\newcommand{\unit}[1]{~\text{#1}}

\makeatletter
\gdef\@fpheader{}
\makeatother

\def\sol{\rm sol}
\def\atm{\rm atm}

\def\BR{\text{BR}}

\begin{document}
%\setlength{\baselineskip}{18pt}
%\onehalfspacing 
%%%%%%%%%%%%%%%%%%%%%%%%%%%%%%%%%%%%%%%%%%%%

\title{Constraining the new contributions to electron $g-2$ in a radiative neutrino mass model} 

\author{Bayu Dirgantara}
\email{bayudirgantara@unhas.ac.id}
\affiliation{Department of Physics, Hasanuddin University, Makassar 90245, Indonesia}
\affiliation{School of Physics and Center of Excellence in High Energy Physics and Astrophysics, Suranaree University of Technology, Nakhon Ratchasima 30000, Thailand}

\author{J. Julio}
\email{julio@brin.go.id}
\affiliation{National Research and Innovation Agency, Kawasan Sains dan Teknologi Bacharuddin Jusuf Habibie, South Tangerang 15314, Indonesia}

\begin{abstract}
We examine electron and muon anomalous magnetic dipole moments within a radiative neutrino mass model featuring TeV-scale scalar leptoquarks $S(3,1,-1/3)$ and $R(3,2,1/6)$.  We utilize textures with decoupling electron and muon sectors, so that both electron and muon anomalous magnetic dipole moments could  receive internal chiral enhancements from different heavy up-type quarks while in the same time evading the stringent $\mu\to e\gamma$ constraint. A successful fit to neutrino oscillation data requires the simultaneous presence of one- and two-loop neutrino mass contributions. This severely constrains the parameter space of the model, which results in a negligible new physics correction to the muon $g-2$. The electron $g-2$ discrepancy implied by the rubidium experiment, on the other hand, can be resolved within $2\sigma$ uncertainty provided that neutrino mass ordering is inverted. Lepton-flavor-violating tau decay rates, such as $\tau\to e\gamma$ and $\tau\to 3e$, are predicted to be within the sensitivities of next-generation experiments.
\end{abstract}

\maketitle
%%%%%%%%%%%%%%%%%%%%%%%%%%%%%%%%%%%%%%%%%%%%%%%
%%%%%%%%%%%%%%%%%%%%%%%%%%%%%%%%%%%%%%%%%%%%%%%
\section{Introduction}
The massive nature of neutrinos remains the only direct experimental evidence for physics beyond the Standard Model (SM). However, the dynamical origin of neutrino masses is still unknown and has been extensively studied even before the discovery of neutrino oscillations. While the canonical seesaw mechanism~\cite{Minkowski:1977sc,*Yanagida:1979as,*Gell-Mann:1979vob,*Mohapatra:1979ia,*Glashow:1979nm} provides a compelling explanation, it postulates new physics at a very high energy scale (i.e., of the order $10^{14}~\text{GeV}$), far beyond any foreseeable experimental reach. An alternative and more testable approach is found in radiative neutrino mass models, in which neutrino masses are induced by higher-order quantum loop corrections at a lower energy scale~\cite{Zee:1980ai,Cheng:1980qt,Zee:1985id,*Babu:1988ki}.  These models often predict observable rates for lepton-flavor-violating decays that can be probed by next-generation experiments, providing an independent cross check. See Refs.~\cite{Cai:2017jrq,Babu:2019mfe} for reviews on this subject. 

Besides neutrino masses, other observational data can be employed to test the SM. These include precision measurements of leptonic anomalous magnetic dipole moments (from now on called lepton $g-2$), denoted as $a_\ell\equiv \tfrac{1}{2}(g-2)_\ell$. For muon $g-2$, FNAL E989 has reported an unprecedented accuracy at the level of 127 ppb~\cite{Muong-2:2025xyk,Muong-2:2023cdq,*Muong-2:2021ojo}, confirming the earlier BNL E821 result~\cite{Muong-2:2006rrc}. Interpretation of this result requires a firm SM prediction. The determination of the SM value, however, is now ambiguous due to the tension between data-driven and lattice QCD determinations of the hadronic vacuum polarization contribution. While the latter has been solidified by various groups over the last few years~\cite{RBC:2018dos,*Giusti:2019xct,*Borsanyi:2020mff,*Lehner:2020crt,*Wang:2022lkq,*Aubin:2022hgm,*Ce:2022kxy,*ExtendedTwistedMass:2022jpw,*RBC:2023pvn,*Kuberski:2024bcj,*Boccaletti:2024guq,*Spiegel:2024dec,*RBC:2024fic,*Djukanovic:2024cmq,*ExtendedTwistedMass:2024nyi,*MILC:2024ryz,*FermilabLatticeHPQCD:2024ppc}, the former faces challenges because of the disagreement between recent $e^+e^-\to\pi^+\pi^-$ data from the CMD-3 Collaboration~\cite{CMD-3:2023alj,*CMD-3:2023rfe} and the earlier data. In the 2025 update, the Theory Initiative (TI) Group adopts the lattice result to determine the SM prediction. This results in $\delta a_\mu\equiv a_\mu^{\text{exp}}-a_\mu^{\text{SM}}=38(63)\times 10^{-11}$~\cite{Aliberti:2025beg}, which indicates no statistically significant evidence for new physics.  

Similar situation is found in the electron $g-2$. While the experimental value of this quantity has been significantly improved~\cite{Fan:2022eto} from the previous data~\cite{Hanneke:2008tm}, this has been overshadowed by the determination of the SM value, hindering a clear interpretation. This is due to the $5\sigma$ difference in the measurements of the QED fine-structure constant $\alpha_{em}$, carried out in cesium (Cs) and rubidium (Rb) atoms~\cite{Parker:2018vye,Morel:2020dww}. These measurements lead to conflicting deviations: $\delta a_e=(-10\pm2.6)\times 10^{-13}$ in Cs and $\delta a_e=(3.5\pm1.6)\times 10^{-13}$ in Rb, corresponding to $-3.8\sigma$  and $+2.2\sigma$ discrepancies, respectively. Although there may be still a room for a new physics, the inconsistency requires further investigation, and at present, we can only treat them as two independent results. In this context, it would be interesting to find models that provide a unified explanation for neutrino masses and the observed values of lepton $g-2$, particularly those that can have preference over any discrepancy in the electron $g-2$.

Motivated by that, we select a model featuring two scalar leptoquarks $S(3,1,-1/3)$ and $R(3,2,1/6)$. The leptoquark $S$ is known to induce large corrections to lepton $g-2$ via internal chiral enhancement, as it can couple to both left-handed and right-handed of the up-type quarks and charged leptons~\cite{Leveille:1977rc,*Djouadi:1989md,*Davidson:1993qk,*Couture:1995he,*Chakraverty:2001yg,*Biggio:2014ela,*Bauer:2015knc,*Das:2016vkr,*ColuccioLeskow:2016dox}. Moreover, the leading contributions to neutrino masses arise at both one- and two-loop levels. Previous studies on this model, however, focused mostly only on one-loop contributions~\cite{Chua:1999si,Mahanta:1999xd,Dorsner:2017wwn,Cai:2017wry,Zhang:2021dgl,Parashar:2022wrd,Dev:2024tto}. The two-loop analysis was given in Ref.~\cite{Babu:2010vp}, but it did not include one-loop contributions. To the best of our knowledge, no previous study has simultaneously incorporated both one- and two-loop contributions to neutrino mass within this specific model while also addressing the current lepton $g-2$ situation.\footnote{For other neutrino mass models attempting to address these anomalies, see~\cite{Abdullah:2019ofw,CarcamoHernandez:2020pxw,Chen:2020jvl,Dutta:2020scq,Arbelaez:2020rbq,Jana:2020joi,DelleRose:2020oaa,Cao:2021lmj,Mondal:2021vou,Hernandez:2021kju,Escribano:2021css,De:2021crr,Borah:2021khc,Barman:2021xeq,Chowdhury:2022jde,Julio:2022ton,Julio:2022bue,Primulando:2022vip,Thao:2023gvs}.}

The objective of this paper is twofold. First, we demonstrate how one- and two-loop terms of neutrino masses are equally important to fit neutrino oscillation data. Second, based on the fit, we determine the corrections to lepton $g-2$. To achieve this goal, we employ particular textures---as discussed in Refs.~\cite{Bigaran:2020jil,Dorsner:2020aaz}---where the electron and muon sectors are effectively decoupled. This configuration ensures that the respective anomalous magnetic moments are induced by chiral enhancements from different up-type quarks, while simultaneously avoiding the potentially dangerous chirally-enhanced $\mu\to e\gamma$ decay. Applying  these textures to the neutrino mass formula given in this model, we find that the lightest neutrino is massless although both normal and inverted neutrino mass orderings can be admitted. Owing to the simultaneous presence of one- and two-loop neutrino mass terms, neutrino data strongly constrain the parameter space of the model, leading to a small new physics correction to the muon $g-2$. Furthermore, only the electron $g-2$ discrepancy implied by the Rb measurement can be resolved within $2\sigma$ uncertainty, which only occurs in the inverted mass ordering. Finally, rates for LFV tau decays, such as $\tau\to3e$ and $\tau\to e\gamma$, are found to be near the current limits, and are thus accessible to future experiments. 

The rest of the paper is organized as follows. In Sect.~\ref{sec:Model}, we briefly discuss the model. The expression for neutrino mass formula and the choice of Yukawa textures as well as their implications on neutrino mass generation are discussed in Sect.~\ref{sec:numass}.  Relevant constraints are presented in Sect.~\ref{sec:exp-cons}, before we give our conclusions in Sect.~\ref{sec:concl}.

%%%%%%%%%%%%%%%%%%%%%%%%%%%%%%%%%%%%%%%%%%%%%%%
\section{Brief review of the model}\label{sec:Model}
%%%%%%%%%%%%%%%%%%%%%%%%%%%%%%%%%%%%%%%%%%%%%%%

The model we consider consists of two leptoquarks $S(3,1,-1/3)$ and $R(3,2,1/6)\equiv(R^{2/3},R^{-1/3})^T$. Together with the SM particles, they form new Yukawa interactions as follows
\begin{align}
	\mathcal{L}_{Y}^{\rm new} = ~& \lambda_{ij}^{L}Q^{T}_{i}C\epsilon L_{j}S^\dagger + \lambda_{ij}^{R}u^{T}_{Ri}C e_{Rj}S^\dagger +\lambda_{ij}\bar d_{Ri}R^T\epsilon L_{i} 
	+ \hc 
 \label{eq:lq}
\end{align}
Here, $i,j=1-3$ are generation indices, $C$ is the charge conjugation operator, and $\epsilon$ is the $SU(2)$ antisymmetric tensor. All terms in \eq{eq:lq} are capable of inducing corrections to the lepton $g-2$. Interactions mediated by the $S$ leptoquark are of particular interest because they can induce a large correction via internal chirality flip ($S$ couples to both left-handed and right-handed up-type quarks), especially when we have the top quark inside the loop.

It should be noted that, besides being gauge invariant, all terms in \eq{eq:lq} also respect baryon number ($B$) and lepton number ($L$) symmetries. Thus, diquark terms, $QQS$ and $u_Rd_RS$, are not allowed since both break the $B$ symmetry. Although it may sound {\it ad hoc}, such a symmetry is needed to avoid the proton decay. As for the lepton number, it can only be broken by a trilinear coupling appearing in the scalar potential. That is,
\begin{align}
    V=&~ \mu^2_R R^\dagger R + \mu^2_S S^\dagger S + \lambda' (H^\dagger H)^2 + \lambda'' (R^\dagger R)^2 + \lambda''' (S^\dagger S)^2 + \tilde{\lambda}_{HR} (H^\dagger R)(R^\dagger H) + \lambda_{HR} (H^\dagger H)(R^\dagger R) \nonumber \\
    &+\lambda_{HS} (H^\dagger H)(S^\dagger S) + \lambda_{RS} (R^\dagger R)(S^\dagger S) + (\mu R^\dagger H S + \hc) 
    \label{eq:sc-pot}
\end{align}
Note that the parameter $\mu$ is taken to be real since its complex phase can always be rotated away.
With all Yukawa coupling matrices of Eq.~\eqref{eq:lq} and the dimensionful coupling $\mu$ of Eq.~\eqref{eq:sc-pot} present, the lepton number is no longer the symmetry of the model.  An alternative way of seeing this is by integrating out the heavy leptoquark states. This will result in two  $\Delta L=2$ operators~\cite{Babu:2001ex,*deGouvea:2007qla,*Bonnet:2009ej,*Angel:2012ug}, i.e., (i) $(\bar d_RL\epsilon H)(L\epsilon Q)$ and (ii) $(\bar d_RL\epsilon H)(u_Re_R)$, indicating that, at leading order,  neutrino masses will be generated in two different ways. The operator (i), for instance, will generate neutrino masses at the one-loop level after one connects the $d_R$ and $d_L$ legs. Similarly, the operator (ii), with the help of the SM charged current, will induce neutrino masses at the two-loop level. Note that the two-loop contribution is not the higher-order correction to the one-loop one since they come from different sets of couplings.

A nonzero $\mu$ of \eq{eq:sc-pot} will induce mixing between $S^{1/3}$ and $R^{1/3}$ leptoquarks
\begin{align}
   \mathcal{L} \supset  \begin{pmatrix} 
    S^{-1/3},  & R^{-1/3}
    \end{pmatrix}
    \begin{pmatrix}
        m_S^2 & \mu v_{EW}/\sqrt{2} \\
        \mu v_{EW}/\sqrt{2} & m_R^2 
    \end{pmatrix}
    \begin{pmatrix}
        S^{1/3} \\ R^{1/3} 
    \end{pmatrix},
    \label{eq:LQ-mass-matrix}
\end{align}
where $v_{EW}=246$~GeV is the electroweak vacuum expectation value, while $m_S^2$ and $m_R^2$ include leptoquark bare masses as well as any possible contributions arising from quartic interactions
\begin{align}
    m_S^2 =~& \mu_S^2 + \tfrac{1}{2}\lambda_{HS}v_{EW}^2, \nonumber \\
    m_R^2 =~& \mu_R^2 + \tfrac{1}{2}(\lambda_{HR}+\tilde{\lambda}_{HR})v_{EW}^2.
\end{align}
The mass matrix in Eq.~\eqref{eq:LQ-mass-matrix} can be diagonalized by rotating $S^{1/3}$ and $R^{2/3}$ to the mass eigenstates
\begin{align}
    \begin{pmatrix}
        X_1^{1/3} \\ X_2^{1/3} 
    \end{pmatrix}
    =
    \begin{pmatrix}
        c_\theta & s_\theta \\
        -s_\theta & c_\theta
    \end{pmatrix}
    \begin{pmatrix}
        S^{1/3} \\ R^{1/3} 
    \end{pmatrix},
    \label{eq:LQ-mixing}
\end{align}
where $c_\theta,s_\theta$ denote $\cos \theta,\sin \theta$, with $\tan2\theta=\sqrt{2}\mu v_{EW}/(m_S^2-m_R^2)$. The process yields mass eigenvalues for $X_{1,2}^{1/3}$, that is,
\begin{align}
    M_{1,2}^2=\frac{1}{2}\left[ m_S^2+m_R^2\pm\sqrt{(m_S^2-m_R^2)^2+2\mu^2v_{EW}^2}\right].
\end{align}
In general, the mass splitting of LQs will affect the oblique parameters, $S$, $T$, and $U$. We will be mostly concerned with the $T$ parameter, as it is the largest among the three. The correction to this parameter is given by
\begin{eqnarray}
	\Delta T&=&\frac{1}{8\pi m_W^2s_{W}^{2}}\bigg[s_{\theta}^{2}F(M_{1}^{2},M_{3}^{2})+c_{\theta}^{2}F(M_{2}^{2},M_{3}^{2})  -\frac{1}{4}s_{2\theta}F(M_{1}^{2},M_{2}^{2})\bigg],
\end{eqnarray}
where $M_3^2\equiv m_R^2-\tfrac{1}{2}\tilde{\lambda}_{HR}v_{EW}^2$, $m_W$, and $s_W\equiv\sin\theta_W$ denote the mass square of $R^{2/3}$ LQ, the mass of $W$ boson, and weak mixing angle, respectively.  The function $F(x,y)$ is found as
\begin{eqnarray}
	F(x,y)&=&\frac{1}{2}(x+y)-\frac{xy}{x-y}\ln\left(\frac{x}{y}\right). 
\end{eqnarray}
Technically, the three LQ masses cannot be degenerate, or else neutrino mass will be zero (see next section). However, the splitting cannot be arbitrarily large. From electroweak precision data, one gets $\Delta T<0.2$ at 95\% CL~\cite{ParticleDataGroup:2024cfk}. To illustrate, the upper limit can be obtained for $M_1=1500$ GeV with $\Delta M\equiv M_2-M_1=50$ GeV and $M_3=1585$ GeV. 

All Yukawa couplings in \eq{eq:lq} are defined in the basis where the charged lepton mass matrix is  diagonal. By the same token, due to the absence of a right-handed charged current in the model, we can, without loss of generality, adopt the same basis for the right-handed quarks. That is, we take a basis where quark mass matrices, appearing in the Lagrangian as $\mathcal{L} \supset \bar q_L M_q q_R$ ($q=u,d$), are such that $M_q$ are diagonalized by $V_q M_q$, with $V_q$ being unitary matrices rotating the left-handed quark fields: $q_L\to V_q q_L$. The product of the two unitary matrices, $V\equiv V_uV_d^\dagger$, is the one that we identify  as the Cabibbo-Kobayashi-Maskawa (CKM) mixing matrix.

Rotating all fields into their mass eigenstates, \eq{eq:lq} becomes
\begin{align}
    \mathcal{L}_{Y}^{\rm new} = ~& \left(\lambda_{ij}^{u} u^T_{Li}C\ell_{Lj} - \lambda^d_{ij}d^T_{Li}C\nu_{Lj} +  \lambda_{ij}^{R}u^{T}_{Ri}C \ell_{Rj}\right) (c_\theta X_1^{1/3} - s_\theta X_2^{1/3}) +\lambda_{ij}\bar d_{Ri} \ell_{Lj} R^{2/3} \nonumber \\
    & -\lambda_{ij} \bar d_{Ri} \nu_{Lj} (s_\theta X_1^{-1/3} + c_\theta X_2^{-1/3})+ \hc,
    \label{eq:lag-mass}
\end{align}
where $\lambda^L$ has been rotated to flavor basis by $\lambda^{q}=V_{q}^\ast\lambda^L$. In the present work, we do not offer a mechanism of determining $V_{q}$. However, we can always make an ansatz, i.e., by fixing the flavor structure of one (rotated) Yukawa coupling matrix ($\lambda^u$ or $\lambda^d$), and determine the other via $\lambda^u=V^\ast\lambda^d$.

%%%%%%%%%%%%%%%%%%%%%%%%%%%%%%%%%%%%%%%%%%%%%%%%%%

\begin{figure*}[t]
    \centering    
    \includegraphics[width=13cm]{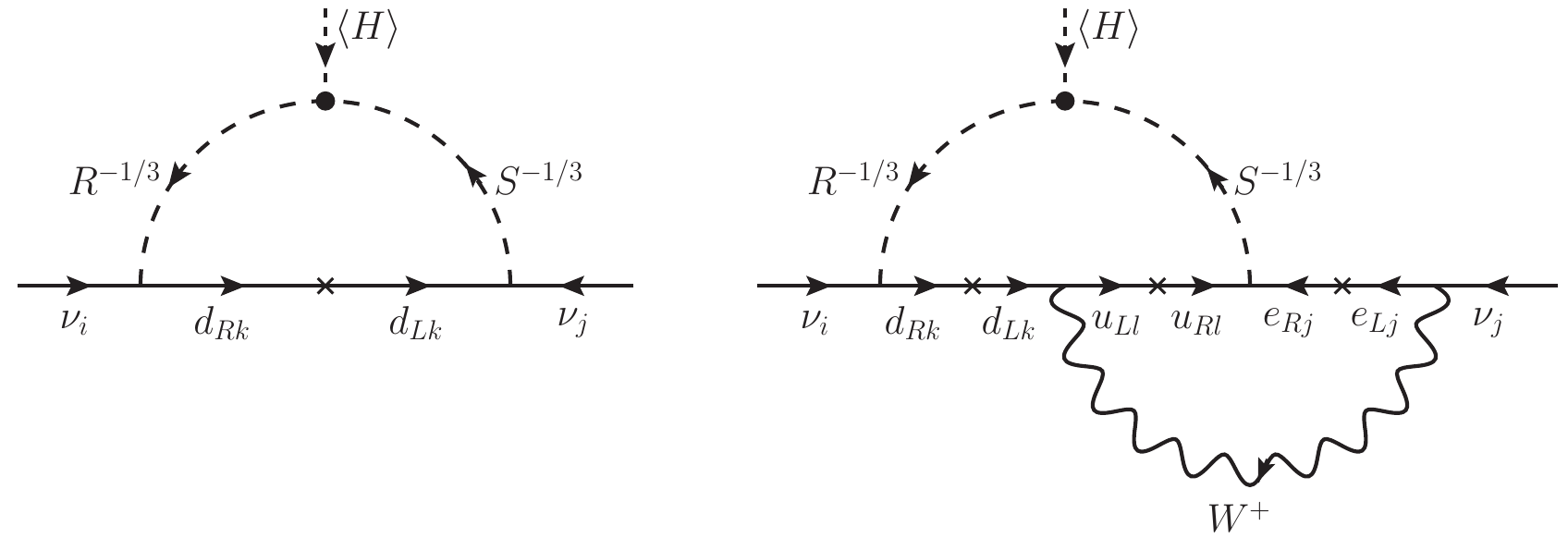}
    \caption{One- and two-loop diagrams leading to neutrino mass generation.}
    \label{fig:Feynman}
\end{figure*}

\section{Neutrino mass generation}
\label{sec:numass}

There are two sets of diagrams that generate neutrino masses, i.e., the one-loop diagrams and two-loop diagrams, see Fig.~\ref{fig:Feynman}. The one-loop diagram involves the exchange of $X_a^{1/3}$ leptoquarks and down-type quarks $d_j$, and it is proportional to the product of $\lambda^L\lambda$ couplings, while the two-loop one involves $X_a^{1/3}$ and $W$ exchanges and is proportional to $\lambda^R\lambda$. The one-loop contributions receive chiral suppression from down-type quark masses, whereas the two-loop contributions receive more chiral suppression factors from all charged fermion masses. Coupled with the additional loop suppression, the two-loop diagrams may only be relevant for relatively narrow range of leptoquark masses, i.e., not more than 10 TeV.  Given that the two sets of diagrams are governed by different sets of couplings, the two-loop diagram cannot be regarded as the higher order corrections to the one-loop one and ought not to be simply omitted.  In fact, there are some flavor textures in which the one- and two-loop diagrams induce comparable strengths.

In evaluating those diagrams, we work in the general 't Hooft $R_\xi$ gauge. Therefore, each propagator of $W$ and the Goldstone boson will contain the gauge parameter $\xi$. However, after summing the two contributions, all $\xi$-dependent terms will add up to zero. The $\xi$-independent terms are the same as those obtained in the unitary gauge. It should also be noted that due to the Majorana property of neutrino mass terms, there are actually other diagrams (not shown in the figure) where one replaces the internal particles with their charged conjugates. Those diagrams are just the transpose of diagrams shown in the Fig.~\ref{fig:Feynman}. Together, they give a flavor-symmetric neutrino mass matrix. 

\begin{figure}[t!]
	\centering
	\subfigure[Without top quark.]{\includegraphics[width=0.45\textwidth]{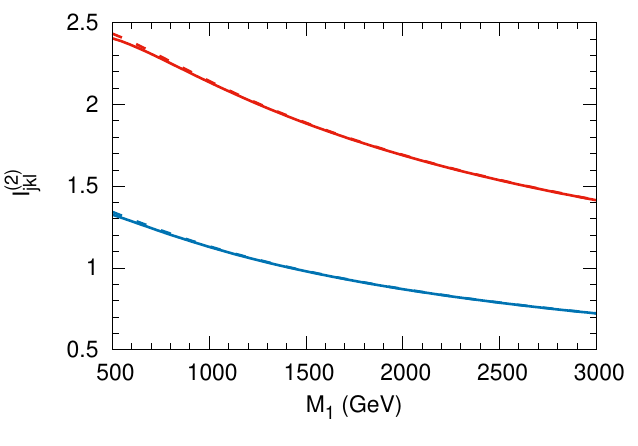}\label{fig:int1}}
	\subfigure[With top quark.]{\includegraphics[width=0.45\textwidth]{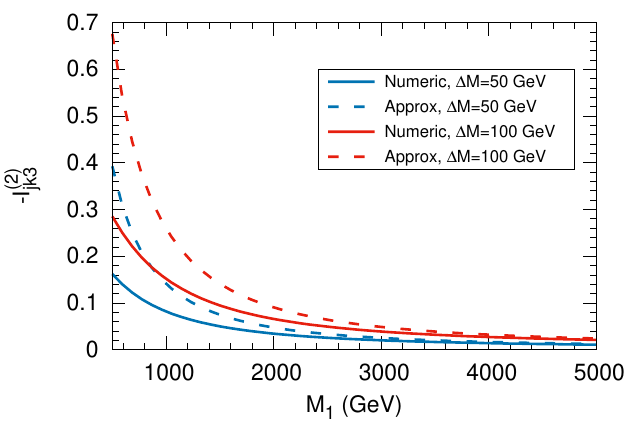}\label{fig:int2}}
	\caption{Comparison between numerical (solid) and approximated (dashed) values of the two-loop integral in the case of light quarks (left) and top quark (right) inside the loop. }
	\label{fig:integral}
\end{figure}

\begin{widetext} 
The neutrino mass matrix is, therefore, given by
\begin{align}
    (M_\nu)_{ji} =  \frac{3s_{2\theta}}{16\pi^2}m_b & \bigg\{ \left[({\lambda^d}^T)_{jk}(D_d)_k\lambda_{ki} + (\lambda^T)_{jk}(D_d)_k\lambda^d_{ki} \right]I_k^{(1)} \nonumber \\
    & + \frac{g^2m_tm_\tau}{16\pi^2 M_1^2} \left[(D_\ell)_j({\lambda^R}^T)_{jl}(D_u)_lV_{lk}(D_d)_k\lambda_{ki} I^{(2)}_{jkl} + (\lambda^T)_{jk}(D_d)_k(V^T)_{kl}(D_u)_l\lambda^R_{li}(D_\ell)_i I^{(2)}_{ikl} \right] \bigg\}.
    \label{eq:numass}
\end{align}
The factor of 3 is the color factor, $g$ is the weak coupling, while $D_d$, $D_u$, and $D_\ell$ are the down-type quark, up-type quark, and charged lepton mass matrices normalized to the third generation masses. The functions, $I_k^{(1)}$ and $I_{jkl}^{(2)}$, are one- and two-loop integral functions,  found as 
\begin{align}
    I_k^{(1)}(M_1^2,M_2^2) = &~\frac{1}{2}\ln\frac{M_1^2}{M_2^2} + \frac{1}{2} m_{d_k}^2\left[\frac{\ln(M_1^2/m_{d_k}^2)}{M_1^2-m_{d_k}^2} - \frac{\ln(M_2^2/m_{d_k}^2)}{M_2^2-m_{d_k}^2} \right], 
    \label{eq:int1}\\
    I_{jkl}^{(2)}(M_1^2,M_2^2,m_W^2) = & \sum_{a=1,2}(-1)^{a+1}\frac{M_{1}^{2}}{M_a^2-m_{d_k}^2}\int_{0}^{1}dx\int_0^\infty dt\;t\left(1+\frac{t}{4m_W^2}\right)\frac{1}{t+m_{\ell_j}^2}\frac{1}{t+m_{W}^2}\nonumber\\
	& \quad \times \ln\left[\frac{x(1-x)t+xm_{u_l}^2+(1-x)M_a^2}{x(1-x)t+x m_{u_l}^2+(1-x)m_{d_k}^2}\right].
 \label{eq:int2}
\end{align} 
Thanks to $m_{d_k}\ll M_{1,2}$, the last term of Eq.~\eqref{eq:int1} is negligible, so the integral is practically flavor-independent. From now on, we will drop the subscript $k$ and simply write the one-loop integral as $I^{(1)}$. The two-loop integral, despite its complicated look, is finite, so it can always be evaluated numerically. We will give approximated expression for the two-loop integral  in two cases: (a)~without top quark  and (b)~with top quark.  

In  case (a), due to $m_{\ell_j,d_{k},u_l}^2\ll m_W^2\ll M_a^2$, one can simply set all the fermion masses to zero to obtain
\begin{align}
    I^{(2)} \equiv I_{jkl}^{(2)}(M_1^2,M_2^2,m_W^2) \simeq & ~\frac{3}{4}\left(1-\frac{M_1^2}{M_{2}^2}\right) \bigg[ 1 + \frac{\pi^2}{3} + 
    f(M_1^2,M_2^2,m_W^2) \bigg],
    \label{eq:appx-light}
\end{align} 
with function $f(a,b,x)$ being defined as
\begin{align}
    f(a,b,x) \equiv~& \frac{1}{b-a}\bigg\{b\ln(a/x)-a\ln(b/x) + \frac{1}{2} \left[b\ln^2(a/x)-a\ln^2(b/x)\right]\bigg\}.
\end{align}
The approximated form works very well, as shown in Fig.~\ref{fig:int1}. Note that, although the $W$ mass is much smaller than LQ masses, it cannot be dropped, or else the integral will be divergent.

In case (b), both top-quark and $W$ masses must be kept. Taking the limit $M_{LQ}\to\infty$, the approximated two-loop integral, at the leading order, is found as
\begin{align}
I^{(2)}_3 \equiv I_{jk3}^{(2)}(M_1^2,M_2^2,m_W^2) \simeq~   \frac{3}{4}\left(1-\frac{M_1^2}{M_{2}^2}\right) &\bigg\{1  + f(M_1^2,M_2^2,m_W^2) \nonumber \\
&+ \left(\frac{m_t^2-m_W^2}{m_W^2}\right)\left[\frac{1}{2}\ln^2\left(\frac{m_t^2-m_W^2}{m_W^2}\right) + {\rm Li}_2\left(\frac{m_W^2}{m_W^2-m_t^2}\right) \right]\nonumber \\
& -\frac{1}{3}\left(\frac{m_t^2}{m_W^2} \right)\left[1 + \frac{\pi^2}{3} + \frac{3}{2}\ln^2\left(\frac{m_t^2}{m_W^2}\right) + f(M_1^2,M_2^2,m_t^2) \right]
\bigg\},
\label{eq:appx-top}
\end{align}
\end{widetext}
where ${\rm Li}_2(x)$ is the dilogarithm function. For a consistency check, one can inspect that Eq.~\eqref{eq:appx-top} will reduce to Eq.~\eqref{eq:appx-light} when $m_t\to 0$. Although the agreement between the numerical value and the approximated form is poor for light LQs, as shown in Fig.~\ref{fig:int2}, it gradually  improves as the LQ masses increase; for instance, when $M_{LQ}\gtrsim 5$ TeV, the difference is already less than 10\%. In the plot, we use the running top mass evaluated at 1 TeV, i.e., $m_t=150$~GeV~\cite{Xing:2007fb,*Babu:2009fd,*Antusch:2025fpm}.

\subsection{Texture choices}  
Following the evaluation of loop integrals, we now address the neutrino mass matrix presented in Eq.~\eqref{eq:numass}. Recall that our motivation is to find a correlation between current observations of the lepton $g-2$ and neutrino oscillation data. To allow a more direct comparison, we consider a scenario with a minimal set of parameters sufficient to induce these phenomena. We require that the corrections to lepton $g-2$ receive chirality enhancements without being precluded by stringent LFV processes, particularly the $\mu\to e\gamma$ bound. With these considerations, we fix the flavor structure of $\lambda^u$, whereas $\lambda^d$ is determined via $\lambda^d=V^T\lambda^u$. This approach is advantageous because it allows us to explicitly prohibit the dangerous $\mu\to e\gamma$ process. Our strategy follows similar logic to the methods used in Refs.~\cite{Cai:2017wry,Babu:2020hun,Julio:2022bue} to mitigate unwanted flavor processes in other neutrino mass models.

We adopt two textures proposed in Ref.~\cite{Bigaran:2020jil} (see also Ref.~\cite{Dorsner:2020aaz}), where the electron and muon are coupled to different massive up-type quarks
\begin{align}
& \text{TX~1}:~ \lambda^u = \begin{pmatrix}
        0 & 0 & 0 \\
        \lambda^u_{21} & 0 & 0 \\
        0 & \lambda^u_{32} & 0 
    \end{pmatrix}, \quad \lambda^R = \begin{pmatrix}
    \ast & \ast & \ast \\
    \lambda^R_{21} & 0 & \ast \\
    0 & \lambda^R_{32} & \lambda^R_{33} 
\end{pmatrix},
\label{eq:tx1} 
\\
& \text{TX~2}:~ \lambda^u = \begin{pmatrix}
        0 & 0 & 0\\ 0 & \lambda^u_{22} & 0 \\ \lambda^u_{31} & 0 & 0
    \end{pmatrix}, \quad \lambda^R = \begin{pmatrix}
    \ast & \ast & \ast \\
    0 & \lambda^R_{22} & \ast \\
    \lambda^R_{31} & 0 & \lambda^R_{33} 
\end{pmatrix}.
    \label{eq:tx2}
\end{align}
In both cases, we assume that the coupling matrix $\lambda$ takes the following form
\begin{align}
	\lambda = \begin{pmatrix}
		0 & 0 & 0 \\
		0 & 0 & 0 \\
		\lambda_{31} & \lambda_{32} & \lambda_{33}
	\end{pmatrix}.
\end{align}
This particular flavor structure ensures that $\lambda$ will not suffer from flavor constraints involving down-type quarks, as these are solely induced by $\lambda^d_{ij}$ and are generally weak, see Sect.~\ref{sec:had}. Interestingly, magnetic dipole terms induced by these particular $\lambda$ couplings are generally suppressed due to the Glashow–Iliopoulos–Maiani (GIM)-like cancellation, see Sect.~\ref{sec:LFV} for a more detailed discussion.

Several remarks are in order. First, the flavor-specific nature of $\lambda^u$ and $\lambda^R$ implies that the corrections to electron and muon $g-2$ can be separately enhanced. In TX~1, for instance, the electron $g-2$ receives an enhancement from the charm mass, while the muon $g-2$ contribution is enhanced by the top mass. Crucially, no dangerous chirally-enhanced $\mu\to e\gamma$ decay is induced, thanks to the vanishing of $(\lambda^u_{22},\lambda^R_{22})$ and $(\lambda^u_{31},\lambda^R_{31})$ pairs. This contrasts with mechanisms where both anomalous magnetic moments are induced by the same up-type quark, as those scenarios face very stringent constraints from the $\mu\to e\gamma$ bound~\cite{Crivellin:2018qmi,Crivellin:2020mjs}.  The only significant constraints come from $K\to\pi\nu\bar{\nu}$ and $Z\to\ell^+\ell^-$ ($\ell=e,\mu$) decays; however, for TeV-scale LQs, these are relatively mild, see Refs.~\cite{Kowalska:2018ulj,Bigaran:2020jil} for a more detailed analysis. Thus, no constraints beyond neutrino data are expected to compromise the outcome of the model. The same reasoning applies to TX~2.

Second, elements of $\lambda^R$  denoted by ``$\ast$'' are not strictly necessary for the present discussion and can be set to zero without affecting the neutrino data fit. This is because the effects of such $\lambda^R_{ij}$ couplings in two-loop neutrino mass generation are suppressed by light fermion masses and CKM matrix elements. Therefore, only terms associated with the top-quark couplings (i.e., $\lambda^R_{3j}$)  are likely to induce significant contributions to the neutrino mass matrix at the leading order. 

Third,  a nonzero $\lambda^R_{33}$ is required to ensure a correct fit to neutrino oscillation data. Its presence in the $(3,3)$ element of the neutrino mass matrix causes the one- and two-loop contributions to appear with comparable strength, leading to the suppression of $\lambda^u$ elements---a key finding of our study. While $\lambda^R_{33}$ may induce LFV tau decays, such decay bounds do not strongly constrain the parameter space of the model, as shown later. An alternative choice, using $\lambda^u_{33}$ in place of $\lambda^R_{33}$, still yields a good fit. However, it causes the one-loop neutrino mass contributions to dominate over the two-loop ones, thereby relaxing the suppression on $\lambda^u$ elements and resulting in a less predictive result.

\subsection{Neutrino mass matrix}
Incorporating the aforementioned textures into the neutrino mass formula presented in Eq.~\eqref{eq:numass} yields a rank-2 neutrino mass matrix, explicitly written as
\begin{align}
    M_\nu = m_0 
    \begin{pmatrix}
        aw & \tfrac{av}{2}+ \tfrac{bw}{2} & \tfrac{a}{2}+\tfrac{w}{2} \\
        \tfrac{av}{2} + \tfrac{bw}{2}  & bv & \tfrac{b}{2} + \tfrac{v}{2} \\
        \tfrac{a}{2}+\tfrac{w}{2} & \tfrac{b}{2} + \tfrac{v}{2} & 1
    \end{pmatrix},
    \label{eq:numass-app1}
\end{align}
where $(a,b,v,w)$ are dimensionless parameters. In TX~1, they are defined as
\begin{align}
    &a = \frac{\lambda_{31}}{\lambda_{33}},~b = \frac{\lambda_{32}}{\lambda_{33}},~v = \frac{1}{{\varepsilon}}\frac{\lambda^u_{32}}{\lambda^R_{33}} + \frac{m_\mu}{m_\tau} \frac{\lambda^R_{32}}{\lambda^R_{33}},~w = \frac{1}{\varepsilon} \frac{V_{cb}}{V_{tb}} \frac{\lambda^u_{21}}{\lambda^R_{33}},
    \label{eq:par1}
\end{align}
while in TX~2 we have
\begin{align}
    v = \frac{1}{\varepsilon} \frac{V_{cb}}{V_{tb}}\frac{\lambda^u_{22}}{\lambda^R_{33}},~~w = \frac{1}{\varepsilon} \frac{\lambda^u_{31}}{\lambda^R_{33}} + \frac{m_e}{m_\tau}\frac{\lambda^R_{31}}{\lambda^R_{33}},
    \label{eq:par2} 
\end{align}
with $a,b$ definitions unchanged. The parameter $m_0=(3s_{2\theta}/8\pi^2)m_b V_{tb}\lambda_{33}\lambda^R_{33}I^{(1)}\varepsilon$ denotes the overall scale of neutrino mass, whereas the parameter $\varepsilon$ quantifies the relative strength of two and one-loop effects and is given by
\begin{align}
    \varepsilon = \frac{g^2m_tm_\tau}{16\pi^2M_1^2}\frac{I^{(2)}_{3}}{I^{(1)}}.
\end{align}
For typical $m_0=0.025$~eV, it must be such that $|\lambda_{33}\lambda^R_{33}s_{2\theta}I^{(1)}\varepsilon|=2.63\times 10^{-10}$. For $M_1=1500$ GeV and $\Delta M=50$ GeV, one gets $I^{(1)}\simeq -0.033$ and  the ratio of the loop integrals is $I^{(2)}_{3}/I^{(1)}\simeq 1.5$. This results in $\varepsilon= 5.6\times 10^{-7}$, leading to $|\lambda_{33}\lambda^R_{33}s_{2\theta}|=0.014$, a value attainable within perturbative limits. This implies that $\lambda^R_{33}$ and $\lambda_{33}$, as well as $s_{2\theta}$, cannot be arbitrarily small to guarantee a correct amount of neutrino masses. Another implication is that $\varepsilon$ should not rapidly drop as LQ masses increase, constraining them to below approximately 10 TeV.

\begin{figure}[t!]
	\centering
	\includegraphics[width=0.75\textwidth]{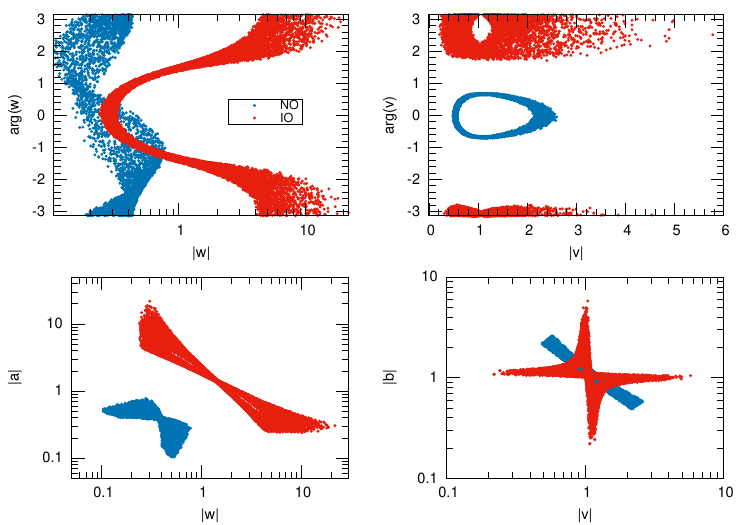}
	\caption{Plots of magnitudes and arguments of neutrino mass parameters in NO and IO. Since the solutions for $a$ ($b$) are complementary to those of $w$ ($v$), only the allowed regions of $w$ and $v$ are shown in the upper panel. The relative strengths between $(a,w)$ and $(b,v)$ are shown in the lower panel.}
	\label{fig:par}
\end{figure}

Other $\lambda^R_{ij}$ couplings (excluding pairs of $(\lambda^R_{31},\lambda^R_{22})$ in TX~1 and $(\lambda^R_{21},\lambda^R_{32})$ in TX~2, which are zero) are not shown because their effects on neutrino mass matrix are negligible. For instance, $\lambda^R_{21}$, which is one of our couplings of interest in TX~1, induces a correction to parameter $w$: 
\begin{align}
    \delta w = \frac{m_c}{m_t}\frac{m_e}{m_\tau}\frac{V_{cb}}{V_{tb}}\frac{I^{(2)}}{I^{(2)}_{3}} \frac{\lambda^R_{21}}{\lambda^R_{33}}.
\end{align}
Using the same LQ mass configuration, such a correction is approximately $\delta w=2.4\times 10^{-6}(\lambda^R_{21}/\lambda^R_{33})$. Since $\lambda^R_{33}\gtrsim 10^{-2}$, this $\delta w$ contribution is highly suppressed, particularly when considering the value of $w$ inferred from neutrino oscillation data, see Fig.~\ref{fig:par}, leaving the coupling $\lambda^R_{21}$ effectively unconstrained. The rest are not relevant to our discussion, so they are simply set to zero. The situation for $\lambda^R_{22}$ in TX~2 can be explained in an analogous way.

Given its rank-2 nature, $M_\nu$ must possess one zero eigenvalue, implying one massless neutrino. However, based on the structure alone, we cannot distinguish the neutrino mass ordering. Thus, in principle, $M_\nu$ can admit both normal mass ordering (NO) and inverted mass ordering (IO). To ensure consistency with neutrino oscillation data, both scenarios require that none of the diagonal elements of $M_\nu$ vanish. The element $(M_\nu)_{33}=m_0$, proportional to the two-loop integral $I^{(2)}_{jk3}$,  represents a purely two-loop contribution and thus cannot be zero. Similarly, a situation where one-loop contributions to $M_\nu$ were absent would lead to $(M_\nu)_{11}=0$. Such a vanishing $(M_\nu)_{11}$ is inconsistent with the current measurement of $\sin^2\theta_{13}$ if the lightest neutrino is massless. This highlights a critical phenomenological requirement: significant contributions from both one-loop and two-loop diagrams to the elements of $M_\nu$ are simultaneously necessary.

Another implication of being a rank-2 matrix is that $M_\nu$ of Eq.~\eqref{eq:numass-app1} has only four independent matrix elements, which are the same as the number of parameters of $M_\nu$. Thus, those elements can be used to express  $(a,b,v,w)$ in terms of neutrino oscillation parameters, i.e.,
\begin{align}
(a,w) =~& m_{13}\pm \sqrt{m_{13}^2-m_{11}}, \quad (b,v) =m_{23} \mp \frac{m_{12}-m_{13}m_{23}}{\sqrt{m_{13}^2-m_{11}}},% \nonumber \\ v =~& m_{23} \pm \frac{m_{12}-m_{13}m_{23}}{\sqrt{m_{13}^2-m_{11}}}, \quad w = m_{13}\mp \sqrt{m_{13}^2-m_{11}},
\label{eq:parosc}
\end{align}
where  $m_{ij}\equiv (M_\nu)_{ij}/(M_\nu)_{33}$, with $M_\nu=U^\ast M_\nu^\text{diag} U^\dagger$. The matrix $U$ is the Pontecorvo-Maki-Nakagawa-Sakata mixing matrix, parametrized by 3 mixing angles and 1 Dirac $CP$ phase
\begin{align}
\begin{small}
    U = \begin{pmatrix}
        1 & 0 & 0 \\
        0 & c_{23} & s_{23} \\
        0 & -s_{23} & c_{23} 
    \end{pmatrix}
    \begin{pmatrix}
        c_{13} & 0 & s_{13}e^{-i\delta} \\
        0 & 1 & 0 \\
        -s_{13}e^{i\delta} & 0 & c_{13}
    \end{pmatrix}
    \begin{pmatrix}
        c_{12} & s_{12} & 0 \\
        -s_{12} & c_{12} & 0 \\
        0 & 0 & 1 
    \end{pmatrix},
\end{small}
\end{align}
while $M_\nu^\text{diag}$ is the diagonal neutrino mass matrix, which contains one Majorana phase. Depending on the mass ordering, it is given by
\begin{align}
    &\text{NO}:~M_\nu^\text{diag} = \text{diag}(0,\sqrt{\Delta m^2_{\sol}}e^{i\alpha},\sqrt{\Delta m^2_{\atm}}), \nonumber \\
    &\text{IO}:~M_\nu^\text{diag} = \text{diag}(\sqrt{\Delta m^2_{\atm}-\Delta m^2_{\sol}},\sqrt{\Delta m^2_{\atm}}e^{i\alpha},0)
    \label{eq:numass-diag}
\end{align}
where $\Delta m_{\sol}^2$ and $\Delta m_{\atm}^2$ correspondingly denote the solar and atmospheric mass splittings. 

\begin{table}[t!]
    \begin{center}
    \caption{Best-fit values of neutrino oscillation parameters with their $1\sigma$ error, taken from the 2024 NuFIT 6.0 analysis~\cite{nufit60,Esteban:2024eli}.}
	\label{tab:nu-exp}
		\begin{tabular}{c|c|c}
			\hline\hline 
			Parameters & NO & IO \\
			\hline
			$s^2_{12}$ &$0.308^{+0.012}_{-0.011}$ &$0.308^{+0.012}_{-0.011}$\\[0.2em]
			$s^2_{23}$ &$0.470^{+0.017}_{-0.013}$ &$0.550^{+0.012}_{-0.015}$\\[0.2em]
			$s^2_{23}$ (higher minima) & $0.575^{+0.012}_{-0.016}$ & $0.468^{+0.016}_{-0.014}$ \\[0.2em]
			$s^2_{13}$ &$0.02215^{+0.00056}_{-0.00058}$ & $0.02231^{+0.00056}_{-0.00056}$\\[0.2em]
			$\Delta m^2_{\sol}/10^{-5}~\text{eV}^2$ &$7.49^{+0.19}_{-0.19}$  &$7.49^{+0.19}_{-0.19}$   \\
			$\Delta m^2_{\atm}/10^{-3}~\text{eV}^2$ &$2.513^{+0.021}_{-0.019}$  &$2.484^{+0.020}_{-0.020}$ \\
			$\delta/^\text{o}$ & $212^{+26}_{-41}$  & $274^{+22}_{-25}$ \\
			\hline\hline
		\end{tabular}
	\end{center}
\end{table}

All neutrino oscillation parameters are summarized in Table~\ref{tab:nu-exp}. Note that some parameter best-fit values, specifically $\delta$ and $\theta_{23}$, exhibit some dependence on the mass ordering.  For instance, the best-fit value for $\delta$  is consistent with the $CP$-conserving value of $180^\text{o}$ in NO. Conversely, in IO, the best-fit $\delta$ is close to maximal $CP$-violating value of $270^\text{o}$ and disfavors the $CP$-conserving value by more than $3\sigma$.

Furthermore, the exact location for the octant of $\theta_{23}$ is still unresolved. It is true that the best-fit value for $\theta_{23}$ is found in the first octant for NO and the second octant for IO. However, current oscillation data cannot conclusively exclude higher-minimum solutions for $\theta_{23}$ in another octant~\cite{Esteban:2024eli}. Specifically, these occur at $\theta_{23}>45^\text{o}$ for NO  and $\theta_{23}<45^\text{o}$ for IO.

The choice of  $\theta_{23}$ octant is relevant for determining the parameters in Eqs.~\eqref{eq:par1} and \eqref{eq:par2}, as their values depend on it. While it is impractical to show their explicit, lengthy expressions, one can examine their term-by-term behavior by expanding them in powers of a small parameter, such as $s_{13}$. Usually, an expansion up to $\mathcal{O}(s_{13}^2)$ is sufficient to give a convergent result. For example, the leading terms for $(a,w)$ are found to be proportional to $\sec\theta_{23}$ in NO and to $\csc\theta_{23}$ in IO; similar patterns are also observed in higher-order terms. This finding, confirmed numerically, indicates that larger values will be obtained when using higher minima (i.e., $\theta_{23}>45^\text{o}$ for NO and $\theta_{23}<45^\text{o}$ for IO). This is the choice that we adopt in our calculation.

We vary each oscillation parameter within its $2\sigma$ uncertainty. The exception is the Majorana phase $\alpha$, which is allowed to take the entire range of $[0,2\pi]$.  Through Eq.~\eqref{eq:parosc}, parameters $(a,b,v,w)$ are directly determined, and  the results are presented in Fig.~\ref{fig:par}. As expected, most couplings are of $\mathcal{O}(1)$ in both mass orderings. Notable exceptions are parameters  $a$ and $w$, which can reach the magnitudes of $20$ in IO. This value is reached at the $2\sigma$ edge of the $CP$ phase $\delta$. Future improvements that narrow this uncertainty toward the maximal $CP$-violating value will reduce the resulting magnitudes of $a,w$.

By employing Eqs.~\eqref{eq:par1} and \eqref{eq:par2}, we determine their impact on the lepton anomalous magnetic moments. For this purpose, we choose the benchmark LQ masses of $M_1=1500$ GeV and $\Delta M=50$~GeV, together with the maximal LQ mixing ($\theta=\pi/4$). These benchmark values are consistent with current LQ direct searches at LHC~\cite{CMS:2018oaj,*ATLAS:2021oiz,*CMS:2022nty,*ATLAS:2022wcu}.

\subsection{TX~1}
We now turn to discussing the implication on texture TX~1. We start with the correction to muon $g-2$, $\delta a_\mu$. This is primarily governed by $\lambda^u_{32}$ and $\lambda^R_{32}$, which are related to the parameter $v$ through Eq.~\eqref{eq:par1}. Given that $\varepsilon=5.6\times 10^{-7}$ and $\lambda^R_{33}\gtrsim 10^{-2}$, the coupling $\lambda^u_{32}$ is strongly constrained.  Considering the value of $v$, for a typical value of $|\lambda^R_{33}|\sim 1$, we find $|\lambda^u_{32}|\simeq \mathcal{O}(10^{-6})$. In contrast, $\lambda^R_{32}$ and $\lambda^R_{33}$ are only mildly constrained by the $Z\to \mu^+\mu^-$ and $Z\to\tau^+\tau^-$ decays. At $2\sigma$ level, each process places $|\lambda^R_{32}|\lesssim 2.79$ and $|\lambda^R_{33}|\lesssim 1.92$, respectively, see Sect.~\ref{sec:Zll} for a more detailed discussion. However, in reality both couplings are simultaneously present, so we need to consider the constraints from LFV tau decay bounds, such as $\tau\to\mu\gamma$ or $\tau\to 3\mu$.

With such a small $\lambda^u_{32}$, the supposedly top-enhanced term only induces marginal effects on  $\delta a_\mu$.  The leading contribution to this correction comes from the term with external chirality flip, which is proportional to $|\lambda^R_{32}|^2$. Considering the bound on $\lambda^R_{32}$, this implies that $\delta a_\mu$ cannot exceed $11\times 10^{-11}$, which is consistent with the recent result of the TI based on various lattice calculations. Similarly, the suppression of such top-enhanced contribution also weakens LFV tau decay constraints, with the strongest coming from $\tau\to3\mu$ decay. For a fixed $\lambda^R_{33}=1.92$, this yields  $|\lambda^R_{32}|\lesssim 0.58$.

The correction to the electron $g-2$, $\delta a_e$, depends on $\lambda^u_{21}$ and $\lambda^R_{21}$. Again by virtue of Eq.~\eqref{eq:par1}, one can infer $\lambda^u_{21}=w\varepsilon (V_{tb}/V_{cb})\lambda^R_{33}$. The resulting $\lambda^u_{21}$ differs significantly between the two neutrino mass orderings, thanks to the value of $w$. We find
\begin{align}
\text{NO}:~ &1.4\times 10^{-6} \lesssim |\lambda^u_{21}/\lambda^R_{33}| \lesssim 10^{-5} \nonumber \\
\text{IO}:~ &3.2 \times 10^{-6} \lesssim |\lambda^u_{21}/\lambda^R_{33}| \lesssim 2.9 \times 10^{-4},
\end{align}
where the upper bound of the coupling ratio for IO is about one order of magnitude larger than that of NO. Nevertheless, unlike $\delta a_\mu$, the contributions to $\delta a_e$ are still dominated by the internal chirality enhancement via the charm-quark mass. 

Similar to $\lambda^R_{32}$ and $\lambda^R_{33}$ cases, the coupling $\lambda^R_{21}$ is also loosely constrained. The $Z\to e^+e^-$ decay does not impose a significant bound because the amplitude is suppressed by $m_Z^2/M_{LQ}^2$, due to the charm quark loop. Similarly, the $\tau\to \mu e^+e^-$ decay, where $\lambda^R_{21}$ plays a role via box diagrams, can be easily evaded by setting a sufficiently small $\lambda^R_{32}$. Instead, the most significant constraint comes from the high-$p_T$ dilepton search (via $c\bar{c} \to e^+e^-$ channel)~\cite{Greljo:2017vvb}, giving $|\lambda^R_{21}|\lesssim 1.4$ with $36.1~\text{fb}^{-1}$ dataset.

Another important observation is  the electric dipole moment of the electron, discussed in Sect.~\ref{sec:edm}. This mainly constrains the complex phase of $\lambda^R_{21}$, requiring $\arg(\lambda^R_{21})=\arg(\lambda^u_{21})+n\pi$, with $n$ being an unconstrained integer.  As a consequence, $\delta a_e \propto |\lambda^u_{21}\lambda^R_{21}|(-1)^n$, allowing $\delta a_e$ to take positive or negative values. The ability to take both signs is important because it allows the induced value of $\delta a_e$ to be directly checked against results from cesium and rubidium experiments. 

\begin{figure}[t!]
	\centering
	\subfigure[~TX 1]{\includegraphics[width=8cm]{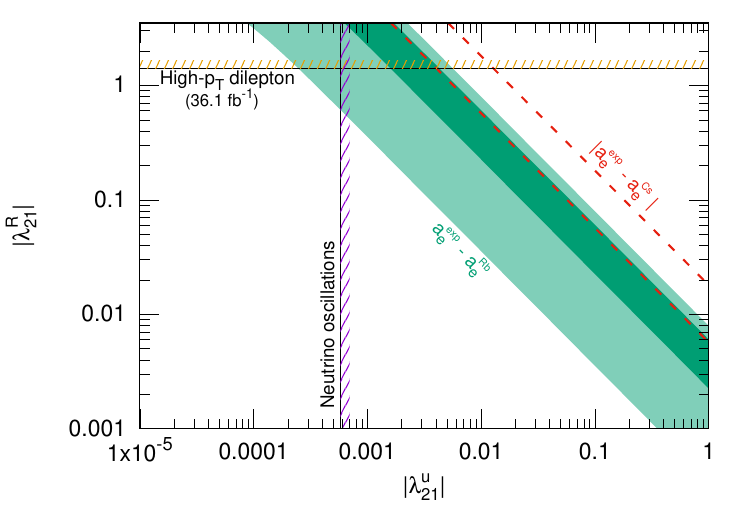}\label{fig:tx1}}
	\subfigure[~TX 2]{\includegraphics[width=8cm]{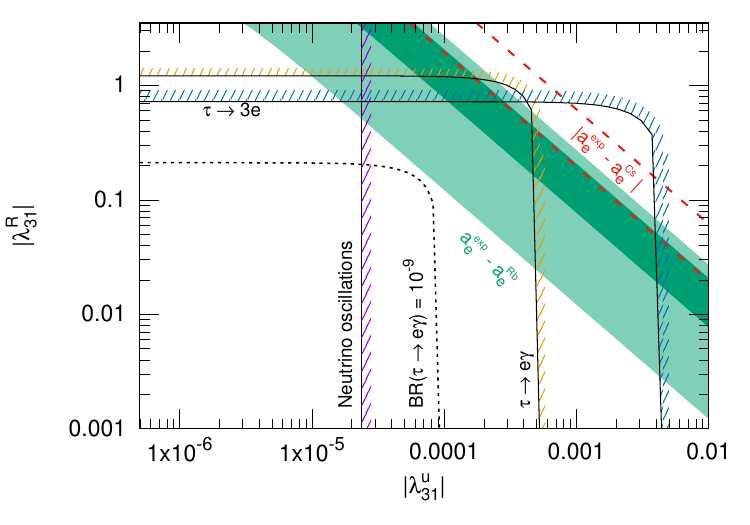}\label{fig:tx2}}
	\caption{The allowed region presented in $|\lambda^u_{21}|$ vs $|\lambda^R_{21}|$ plane for TX~1 (a) and in $|\lambda^u_{31}|$ vs $|\lambda^R_{31}|$ plane for TX~2 (b). The darker (lighter) shaded area corresponds to one- (two-) sigma allowed region of $a_e^\text{exp}-a_e^\text{Rb}$, while the area bordered by the dashed lines correspond to two-sigma allowed region of $|a_e^\text{exp}-a_e^\text{Cs}|$.}
	\label{fig:g-2}
\end{figure}

Fig.~\ref{fig:tx1} illustrates the results of our analysis. The most relevant constraint for each coupling is indicated by the horizontal and vertical lines. Of particular interest is the constraint on $\lambda^u_{21}$, which arises mainly from neutrino oscillation data, with the coupling $\lambda^R_{33}$ being fixed to its maximum value of 1.92.  The bound on $\lambda^u_{21}$ is so strong that it eliminates all but a small region consistent with the $a_e^\text{exp}-a_e^\text{Rb}$ value (shaded area). On top of that, this bound completely excludes the region associated with $|a_e^\text{exp}-a_e^\text{Cs}|$ (bordered by dashed lines), thereby favoring only the NP scenario suggested by the rubidium measurement. Note that this bound is derived based on the results in IO context. For NO, as mentioned before, the bound will be approximately one order of magnitude stronger, ruling out the entire displayed region.

Future experimental results will significantly affect the outcome of this analysis. First, the high-luminosity LHC data ($\sim 3000~\text{fb}^{-1}$) are projected to improve the limit on $|\lambda^R_{21}|$ to approximately $0.8$, further shrinking the allowed region. Second, precise determinations of the neutrino mass ordering, the octant of $\theta_{23}$, and the Dirac $CP$ phase $\delta$ will be crucial for either validating or falsifying the present scenario. For instance, a value of $\delta$ near the maximal $CP$-violating value of $270^\text{o}$, even in IO,  will significantly tighten the upper bound on $\lambda^u_{21}$. Similarly, if NO proves to be the true mass ordering, our scenario would be unable to induce a significant correction to the electron $g-2$. Still, a definitive clarification on the value of the electromagnetic fine-structure constant will be important for a conclusive interpretation of the electron $g-2$ within this framework.

\subsection{TX~2}
The TX~2 scenario is analogous to TX~1. Using $v$ and $w$ values inferred from Fig.~\ref{fig:par}, it is straightforward to determine $\lambda^u$ couplings. Specifically, we find  $|\lambda^u_{22}|= |v\varepsilon(V_{tb}/V_{cb})\lambda^R_{33}|\lesssim 1.6\times 10^{-4}$ and $|\lambda^u_{31}|\simeq |w\varepsilon \lambda^R_{33}|\lesssim 2.1\times 10^{-5}$, assuming IO and the upper value of $\lambda^R_{33}=1.92$ set by the $Z\to \tau^+\tau^-$ decay. On the other hand, the coupling $\lambda^R_{31}$ is practically unconstrained by neutrino data, thanks to the presence of $m_e/m_\tau$ suppression factor in the formula for $w$.  Therefore, in principle, it can take the maximum value allowed by flavor constraints, which are $Z\to e^+e^-$, $\tau\to e\gamma$, $\tau\to 3e$, and $\tau\to e\mu^+\mu^-$ decays. These bounds could be competitive to each other, depending on $\lambda^R_{33}$ and $\lambda^u_{31}$ values. With $\lambda^R_{33}$ set to its maximum value and $|\lambda^u_{31}|\lesssim 2.1 \times 10^{-5}$, it is $\tau\to 3e$  decay that gives the strongest constraint. 

The strong bound on $\lambda^u_{22}$ prevents $\delta a_\mu$ from receiving significant enhancement from the charm-quark mass. Instead,  it is predominantly induced by terms proportional to $|\lambda^R_{22}|^2$. The coupling $\lambda^R_{22}$, like $\lambda^R_{21}$ in TX~1, is constrained by the high-$p_T$ dilepton search via $c\bar{c}\to \mu^+\mu^-$, resulting in  $|\lambda^R_{22}|\lesssim 1.33$. This leads to a predicted $\delta a_\mu\lesssim 2.85\times 10^{-11}$, which is in a good agreement with the TI result. 

The results for this texture are summarized in Fig.~\ref{fig:tx2}. Similar to TX~1, only a small portion of region associated with the rubidium value is allowed, whereas that corresponding to the cesium measurement is entirely ruled out. This reaffirms the significant role of  neutrino data in constraining the allowed parameter space. Future measurements of neutrino oscillation parameters will impact this texture the same way they affect TX~1. 

Interestingly, in the correlation with LFV, the allowed region further suggests that LFV tau decays, such as $\tau\to e\gamma$ and $\tau\to 3e$, should be observed by next-generation experiments. Specifically, we find $\BR(\tau\to e\gamma)\gtrsim 6\times 10^{-9}$ and $\BR(\tau\to 3e)\gtrsim 1.5\times 10^{-8}$, which are within future sensitivity rates, see Table~\ref{tab:LFV}. A failure to observe these LFV decays   within  the projected limits would imply the exclusion of this scenario. In this context, LFV muon decays are suppressed. $\mu\to e\gamma$ suffers from the GIM-like cancellation (see Sect.~\ref{sec:LFV}), while $\mu\to 3e$ and $\mu-e$ conversion in nuclei are suppressed because the corresponding couplings, $\lambda_{31},\lambda_{32}$ (both are correlated with $\lambda_{33}$ via neutrino mass matrix parameters $a$ and $b$) are small due to the assumption of LQ maximal mixing. If the mixing is allowed to be small, $\lambda_{3j}$ couplings could be large, leading to sizable rates for $\mu\to3e$ and $\mu-e$ conversion. We find that $\text{BR}(\mu\to 3e)\lesssim 1.4\times 10^{-15}$ and $R_{\mu e}(\text{Al})\lesssim 3.5\times 10^{-18}$, which are within the sensitivities of next-generation experiments~\cite{Blondel:2013ia,COMET:2018auw,Mu2e:2014fns}.

\subsection{Perturbative cutoff}
Our analysis requires the simultaneous presence of one- and two-loop neutrino mass terms. As mentioned earlier,  the two-loop contributions will maintain their significance as long as the LQ masses do not exceed approximately 10 TeV. Furthermore, some Yukawa couplings are constrained to be of order 1, with $\lambda^R_{33}$ being as large as $1.92$, whereas the others, e.g., $\lambda^u_{ij}$ (or $\lambda^L$ in the original basis) are very restricted. Thus, it is important to verify the ultraviolet (UV) consistency of the model. We evaluate the renormalization group (RG) evolution of these couplings using the following $\beta$ functions
\begin{align}
	16\pi^2 \frac{d\lambda^R_{ij}}{dt} =&~ (Y_u^TY_u^\ast\lambda^R)_{ij} + 2(\lambda^R\lambda^{R\dagger}\lambda^R)_{ij} + \left[2\Tr(\lambda^{L\dagger}\lambda^L) + \Tr(\lambda^{R\dagger}\lambda^R)\right]\lambda^R_{ij} - \left(\frac{13}{3}g'^2 + 4g_3^2\right)\lambda^R_{ij}, \nonumber \\
	16\pi^2 \frac{d\lambda^L_{ij}}{dt} =&~ \frac{1}{2}(Y_u^\ast Y_u^T\lambda^L)_{kj} + 2(\lambda^L\lambda^{L\dagger}\lambda^L)_{ij} + \left[2\Tr(\lambda^{L\dagger}\lambda^L) + \Tr(\lambda^{R\dagger}\lambda^R)\right]\lambda^L_{ij} + 2(Y_u^\ast\lambda^RY_e^\dagger)_{ij} \nonumber \\
	&- \left(\frac{5}{6}g'^2+\frac{9}{2}g^2+4g_3^2\right)\lambda^L_{ij}, \nonumber \\	
	16\pi^2\frac{dy_t}{dt} =&~\frac{9}{2}y_t^3  + \frac{1}{2}(\lambda^R_{33})^2y_t - \left(\frac{17}{12}g'^2+\frac{9}{4}g^2 + 8 g_3^2 \right)y_t, \nonumber \\
	16\pi^2 \frac{dg_i}{dt} =&~ b_ig_i^3.
\end{align}
Here $Y_u$ is the up-type Yukawa coupling matrix, of which only the top Yukawa coupling is kept, while $\{b_1,b_2,b_3\}=\{21/3,-8/3,-13/2\}$ are the coefficients for the three gauge couplings $g',g,g_3$. Other Yukawa couplings  are neglected due to their small values, with the exception of the term proportional to $Y_u\lambda^RY_e^\dagger$ in the beta function of $\lambda^L$.

Numerical evaluation of these RGEs demonstrates that this model is indeed perturbative and technically natural for a broad range of energies. For instance, $\lambda^R_{33}$ will not hit the perturbative limit of $\sqrt{4\pi}$ up to approximately 100~TeV. While this indicates the need for a UV completion at or below this scale, this cutoff is sufficiently high to ensure that the phenomenological results presented here—which are evaluated at the TeV scale—remain robust. In addition, the RGEs show that $\lambda^L_{ij}$  couplings do not receive large corrections across this energy range; their running either is proportional to $\lambda^L_{ij}$ or is driven by the term proportional to $Y_u^\ast\lambda^RY_e^\dagger$, highly suppressed by the small SM Yukawa couplings.

\subsection{Coupled electron and muon sectors}

Before we end this section, we would like to discuss the situation where both $\delta a_e$ and $\delta a_\mu$ are induced via chiral enhancement of the same internal up-type quark, e.g., the top quark. The nonzero elements of $\lambda^u$, i.e., $\lambda^u_{31},\lambda^u_{32}$,  are linked to the neutrino mass matrix parameters through
\begin{align}
    v = \frac{1}{\varepsilon}\frac{\lambda^u_{32}}{\lambda^R_{33}} + \frac{m_\mu}{m_\tau}\frac{\lambda^R_{32}}{\lambda^R_{33}},~~ w = \frac{1}{\varepsilon}\frac{\lambda^u_{31}}{\lambda^R_{33}}+ \frac{m_e}{m_\tau}\frac{\lambda^R_{31}}{\lambda^R_{33}}. 
\end{align}
To address the electron $g-2$, one needs the chiral enhancement, so both $\lambda^R_{31}$ and $\lambda^u_{31}$ must be present. On the contrary, muon $g-2$ does not need a new physics, as implied the current lattice findings. However, the $\mu\to e\gamma$ decay cannot be arbitrarily suppressed because both $\lambda^u_{32}$ and $\lambda^R_{32}$ cannot be simultaneously zero, thanks to $v\simeq 1$ from oscillation data.  To see this more transparently, we express the branching ratio of $\mu\to e\gamma$  in terms of $\delta a_e$ and $\delta a_\mu$  through \cite{Crivellin:2018qmi,Dorsner:2020aaz}
\begin{align}
    \BR(\mu\to e\gamma) = \frac{\tau_\mu\alpha m_\mu^3}{16}\left(\frac{\delta a_e^2}{m_e^2}|x|^2 + \frac{\delta a_\mu^2}{m_\mu^2}\frac{1}{|x|^2} \right),
    \label{eq:bremu}
\end{align}
where $\tau_\mu=2.19\times 10^{-6}~\text{s}$ is the lifetime of the muon and $x\equiv \lambda^u_{32}/\lambda^u_{31}$. Now, let us suppose $\lambda^R_{32}=0$, so $\lambda^u_{32}=v\varepsilon\lambda^R_{33}\lesssim \mathcal{O}(10^{-6})$. Consequently,  $|\delta a_\mu|\ll 10^{-11}$, and thus it can be dropped from Eq.~\eqref{eq:bremu}. Due to $m_e/m_\tau$ factor, $\lambda^R_{31}$ practically has negligible impact on $w$, so it is a good approximation to take $x\simeq v/w$, whose value is between 0.087 and 4.7. With such values, one cannot satisfy the recent bound of $\BR(\mu\to e\gamma)< 1.5\times 10^{-13}$~\cite{MEGII:2025gzr}, unless $\delta a_e\lesssim 10^{-15}$, two orders of magnitude smaller than the recent values implied by both cesium and rubidium experiments. This shows that, even with the muon $g-2$ anomaly gone, the  tension between neutrino data and $\mu\to e\gamma$ bound restricts the electron $g-2$ from getting a significant correction of new physics.

%%%%%%%%%%%%%%%%%%%%%%%%%%%%%%%%%%%%%%%%%%%%%%%%%%
\section{Experimental constraints}
\label{sec:exp-cons}
%%%%%%%%%%%%%%%%%%%%%%%%%%%%%%%%%%%%%%%%%%%%%%
\subsection{Lepton anomalous magnetic moment}
\label{sec:amm} 
The scalar leptoquarks in this model can induce a correction to the anomalous magnetic dipole moment of a lepton, which proceeds via penguin diagrams, in which the photon can be emitted from internal quark and leptoquark lines. Using \eq{eq:lag-mass}, it is found that
\begin{eqnarray}
	\delta a_{\ell} &=& -\frac{3 m_{\ell}^2}{8\pi^{2}}\frac{\zeta_{a}}{M_a^2}\bigg[\left(|\lambda^{u}_{k\ell}|^{2}+\abs{\lambda^{R}_{k\ell}}^{2}\right)\kappa(x_{ka}) + \frac{m_{u_{k}}}{m_\ell}\Re\left(\lambda^{u\ast}_{k\ell}\lambda^{R}_{k\ell}\right)\kappa'(x_{ka})\bigg],
 \label{eq:dal}
\end{eqnarray}
where $\zeta_1(\zeta_2)=c^2_\theta(s^2_\theta)$, $x_{ka}=m_{u_k}^2/M_a^2$. The loop functions $\kappa(x),\kappa'(x)$ are given by
\begin{align}
\kappa(x) =&~ \frac{-1-4x+5x^2-2x(2+x)\ln x}{12(1-x)^4},\nonumber \\
\kappa'(x) =&~ \frac{7-8x+x^2+(4+2x)\ln x}{6(1-x)^3}.
\label{eq:dipole-func}
\end{align}
Note that in \eq{eq:dal} only contributions from $X_a^{1/3}$ and up-type quark exchange are shown. Other contributions involving $R^{2/3}$ and $d_j$ quarks exchange (driven by $\lambda_{ij}$ couplings) are negligible. This can be explained as follows. In the limit of $m_{b}=0$, the loop function of the diagram with photon attached to the down-type quark line is two times larger than the one with photon attached to the leptoquark $R^{2/3}$ line, whereas the $b$-quark electric charge is two times smaller and opposite in sign to that of leptoquark. Adding the two contributions will result in a vanishing amplitude at the leading order, similar to the GIM cancellation. Nonzero terms, i.e., the analog of $\kappa(x)$ function for this LQ exchange, will be proportional to $m_{b}^2/M_{LQ}^2$, which are very suppressed.

The last term of \eq{eq:dal} is induced by an internal quark chirality flip. This is possible because within this model the $X_a^{1/3}$ LQs can couple to charged leptons with different chirality. Thus, it will induce a significant correction if the internal quarks are heavier than external leptons; in the present model this can be achieved by having the top quark or the charm quark inside the loop.

%%%%%%%%%%%%%%%%%%%%%%%%%%%%%%%%%%%%%%%%%%%%%%%%%%%%%%%%
\subsection{Lepton-flavor-violating decays}
\label{sec:LFV}

\subsubsection{$\ell_i\to \ell_j\gamma$ decays}
Lepton-flavor-violating decays in general are expected to occur within this model. The first process we consider is the radiative lepton decay $\ell_i\to\ell_j+\gamma$. Its transition amplitude is driven by the following Lagrangian
\begin{align}
\mathcal{L}^{eff}_{\ell_i \to \ell_j + \gamma} = &~ \frac{m_{\ell_i}}{v_{EW}^2} \bar{\ell}_j \sigma^{\mu\nu}\left(\tfrac{e}{2}A_{2R} P_R + \tfrac{e}{2}A_{2L} P_L \right)\ell_i F_{\mu\nu} + \hc,
\label{eq:muegamma}
\end{align}
where $F_{\mu\nu}=\partial_\mu A_\nu-\partial_\nu A_\mu$ is the electromagnetic strength tensor. The branching ratio is determined as
\begin{align}
	\BR(\ell_i \to \ell_j \gamma) =&~ \frac{\tau_{\ell_i}\alpha m_{\ell_i}^5}{4v_{EW}^4} \left(|A_{2L}|^2+|A_{2R}|^2 \right),
    \label{eq:tauegamma}
\end{align}
with $\tau_{\ell_i}$ denoting the lifetime of the decaying particle $\ell_i$. 

At the lowest order, this process arises through one-loop penguin-type diagrams exchanging $X_a^{1/3}$ LQs. The Wilson coefficients $A_{2R,2L}$ are found as
\begin{align}
    A_{2R} =&~ \frac{3}{16\pi^{2}}\frac{\zeta_{a}v_{EW}^2}{M_a^2}  \bigg[\left(\lambda^u_{ki}\lambda^{u\ast}_{kj} +  \frac{m_{\ell_j}}{m_{\ell_i}}\lambda^{R}_{ki}\lambda^{R\ast}_{kj} \right)\kappa(x_{ka}) + \frac{m_t}{m_{\ell_i}} \lambda^{R}_{3i}\lambda^{u\ast}_{3j} \kappa'(x_{ka}) \bigg], \nonumber \\
    A_{2L} =&~ (\lambda^u \leftrightarrow \lambda^R),
\end{align}
As with the $\delta a_\ell$ case, contributions induced by $\lambda_{ij}$ are not included because they suffer from the GIM-like cancellation. Consequently, the $\mu\to e\gamma$ rate, which can only be driven by $\lambda$ couplings here, is found several orders of magnitude lower than the current bound. Meanwhile, considering $\lambda^u$ structures given in Eqs.~\eqref{eq:tx1} and \eqref{eq:tx2}, only its element involving the top quark (i.e., $\lambda^u_{32}$ in TX~1 and $\lambda^u_{31}$ in TX~2) generates $\tau\to \mu/e+\gamma$ decays through internal chirality flip. Even so, due to neutrino data constraint, such a term does not significantly affect the parameter space of the model. The summary of relevant LFV decays is given in Table~\ref{tab:LFV}.

\begin{table}
	\begin{center}
		\caption{Current experimental upper bounds on branching ratios of LFV decays relevant for the present discussion.}
		\label{tab:LFV}
	\begin{tabular}{l|c|c}
		\hline\hline
		Process & Present bound & Future bound\\
		\hline
		$\mu\to e\gamma$ & $1.5\times 10^{-13}$~\cite{MEGII:2025gzr} & $6\times 10^{-14}$~\cite{MEGII:2018kmf} \\
		$\tau\to e\gamma$ & $3.3\times 10^{-8}$~\cite{BaBar:2009hkt} & $\sim10^{-9}$~\cite{Aushev:2010bq} \\
		$\tau\to \mu\gamma$ & $4.2\times 10^{-8}$~\cite{Belle:2021ysv} & $\sim 10^{-9}$~\cite{Aushev:2010bq} \\
		$\mu\to 3e$ & $1.0\times 10^{-12}$~\cite{SINDRUM:1987nra} & $\sim 10^{-16}$~\cite{Blondel:2013ia} \\
		$\tau\to 3e$ & $2.7\times 10^{-8}$~\cite{Hayasaka:2010np} & $\sim 10^{-9}$~\cite{Aushev:2010bq} \\
		$\tau\to 3\mu$ & $2.1\times 10^{-8}$~\cite{Hayasaka:2010np}  & $\sim 10^{-9}$~\cite{Aushev:2010bq} \\
		$\tau\to \mu e^+e^-$ & $1.8\times 10^{-8}$~\cite{Hayasaka:2010np}  & $\sim 10^{-9}$~\cite{Aushev:2010bq} \\
		$\tau\to e\mu^+\mu^-$ & $2.7\times 10^{-8}$~\cite{Hayasaka:2010np}  & $\sim 10^{-9}$~\cite{Aushev:2010bq} \\
		\hline\hline 
	\end{tabular}
	\end{center}
\end{table}

\subsubsection{$\ell_i\to3\ell_j$ decays}

In addition to $\ell_i\to\ell_j\gamma$, LFV decays can also occur in the form of $\ell_i\to3\ell_j$. We will focus on decays to lighter leptons of the same flavor. The diagrams of these processes occur at one-loop level via penguin-type diagrams exchanging photon, the $Z$ boson, and the Higgs boson, together with the box diagrams. However, since the penguin Higgs exchange diagrams are proportional to outgoing lepton masses, they are practically negligible and are not considered here.  Following the notation of \cite{Kuno:1999jp}, the general form of the effective Lagrangian is given by
\begin{widetext}
\begin{align}
    \mathcal{L}^{eff}_{\ell_i\to3\ell_j} =  \frac{1}{v_{EW}^2} \bigg[&m_{\ell_i} \bar{\ell}_j \sigma^{\mu\nu}\left(\tfrac{e}{2}A_{2L} P_R + \tfrac{e}{2}A_{2R} P_L \right)\ell_i F_{\mu\nu} \nonumber \\
     & + g_1 (\bar\ell_j P_R \ell_i)(\bar\ell_j P_R \ell_j) + g_2 (\bar\ell_j P_L \ell_i)(\bar\ell_j P_L \ell_j)
     \nonumber \\
     & + g_3 (\bar\ell_j \gamma^\mu P_R \ell_i)(\bar\ell_j \gamma_\mu P_R \ell_j) + g_4 (\bar\ell_j \gamma^\mu P_L \ell_i)(\bar\ell_j \gamma_\mu P_L \ell_j) \nonumber \\
    & + g_5 (\bar\ell_j \gamma^\mu P_R \ell_i)(\bar\ell_j \gamma_\mu P_L \ell_j) + g_6 (\bar\ell_j \gamma^\mu P_L \ell_i)(\bar\ell_j \gamma_\mu P_R \ell_j) + \hc \bigg],
\end{align}
from which one can write the branching ratio~\cite{Kuno:1999jp}:
\begin{align}
	\BR(\ell_i\to3\ell_j)
    =\frac{\tau_{\ell_i}m_{\ell_i}^{5}}{512\pi^{3}v_{EW}^4}&\left[\frac{1}{24}\left(\abs{g_{1}}^{2}+\abs{g_{2}}^{2}\right)+\frac{2}{3}\left(\abs{g_{3}}^{2}+\abs{g_{4}}^{2}\right)+\frac{1}{3}\left(\abs{g_{5}}^{2}+\abs{g_{6}}^{2}\right)\right.\nonumber\\
	&\left.+\frac{8}{3}\left(\ln\frac{m^2_{\ell_i}}{m^2_{\ell_j}}-\frac{11}{4}\right)\left(\abs{e^2A_{2R}}^{2}+\abs{e^2A_{2L}}^{2}\right) - \frac{8}{3}\Re\left(e^2A_{2R}g_{4}^{*}+e^2A_{2L}g_{3}^{*}\right)\right.\nonumber\\
	&\left.-\frac{4}{3}\Re\left(e^2A_{2R}g_{6}^{*}+e^2A_{2L}g_{5}^{*}\right)\right].
	\label{eq:eff-3l}
\end{align}
\end{widetext}
We ignore contributions of $\lambda^u_{ij}$, except that inducing chirality flip in the dipole coefficient $A_{2R}$. Effective couplings $g_i~(i=1,...,6)$, therefore, are given as
\begin{align}
	g_{1}= &~0,~
	g_{2}=0, \nonumber \\
    g_{3}=&~e^{2}\left(A_{1R}+Z_{R}g^{\ell}_{R} + \tfrac{1}{2}B_{R}\right), \nonumber \\
	g_{4}= &~e^{2}\left(A_{1L}+Z_{L}g^{\ell}_{L} + \tfrac{1}{2}B_{L}\right), \nonumber \\
	g_{5}= &~e^{2}\left(A_{1R}+Z_{R}g^{\ell}_{L}\right), \nonumber \\
	g_{6}=&~e^{2}\left(A_{1L}+Z_{L}g^{\ell}_{R}\right),
    \label{eq:tau3e}
\end{align}
where $A_{1L,1R}$, $Z_{L,R}$, and $B_{L,R}$ denote contributions from photon, $Z$, and box diagrams. They are found as
\begin{align}
    A_{1R} =&~ \frac{3}{16\pi^2} \frac{\zeta_a v_{EW}^2}{M_a^2}\lambda^{R}_{3i}\lambda^{R\ast}_{3j}\tilde\kappa_R(x_{ta}), \nonumber \\
    A_{1L} =&~ \frac{3}{16\pi^2} \frac{v_{EW}^2}{M_3^2} \lambda_{3i}\lambda^\ast_{3j} \tilde\kappa_L(x_{b}), \nonumber \\
    Z_R =& \frac{3}{16\pi^2} \frac{\zeta_a v_{EW}^2}{m_Z^2c_W^2 s_W^2}\lambda^{R}_{3i}\lambda^{R\ast}_{3j}(g_R^{u^c}-g_L^{u^c})F(x_{ta}), \nonumber \\
    Z_L =&~ \frac{3}{16\pi^2} \frac{v_{EW}^2}{m_Z^2c_W^2 s_W^2}\lambda_{3i}\lambda^{\ast}_{3j}(g_L^{d}-g_R^{d})F(x_b), 
    \nonumber \\
    e^2B_{R} =&~ \frac{3}{16\pi^2} \frac{\zeta_a\zeta_b v_{EW}^2}{M_a^2} \lambda^R_{3i}\lambda^{R\ast}_{3j}|\lambda^R_{3j}|^2 b(x_{ta},r_{ba}), \nonumber \\
    e^2B_{L} =& \frac{3}{16\pi^2} \frac{v_{EW}^2}{M_3^2} \lambda_{3i}\lambda^{\ast}_{3j}|\lambda_{3j}|^2b(0,1),
    \label{eq:wc}
\end{align}
with $x_{b}=m_b^2/M_3^2$, $r_{ba}=M_b^2/M_a^2$, and  $g_{R,L}^{f^c}=-g_{L,R}^{f,SM}=-[T_3(f_{L,R})-Q_fs_W^2]$.\footnote{One should not be confused with the notations $m_b$ and $M_b$; the former refers to the bottom-quark mass, while the latter to the mass of~$X_b^{1/3}$.}  Functions $\tilde{\kappa}_{L,R}(x)$, $F(x)$, and $b(x,y)$ are given by
\begin{align}
    & \tilde\kappa_R(x) = \frac{-10+27x-18x^2+x^3+2(-4+6x+x^3)\ln x}{36(1-x)^4},\nonumber \\
    & \tilde\kappa_L(x) = \frac{-4+9x-5x^3+(-4+6x+4x^3)\ln x}{36(1-x)^4},\nonumber \\
    & F(x)=\frac{x(1-x+\ln x)}{(1-x)^2},\nonumber \\
    & b(x,y) = -\tfrac{1}{2}\int_0^\infty dt \frac{t^2}{(t+x)^2(t+y)(t+1)}.
\end{align}

Besides $\ell_i\to3\ell_j$ decays, there are also $\ell_i\to\ell_j\ell^+_k\ell^-_k$ ($j\neq k$) decays. This type of processes is similar to $\ell_i\to3\ell_j$ decays, but the flavor of $\ell_k^+\ell_k^-$ pair is different  from the other outgoing particle. The branching ratio formula is then given as~\cite{Abada:2014kba}
\begin{align}
	\BR(\ell_i\to\ell_j\ell^+_k\ell^-_k)
	=\frac{\tau_{\ell_i}m_{\ell_i}^{5}}{512\pi^{3}v_{EW}^4}&\left[\frac{1}{12}\left(\abs{g_{1}}^{2}+\abs{g_{2}}^{2}\right)+\frac{1}{3}\left(\abs{g_{3}}^{2}+\abs{g_{4}}^{2}\right)+\frac{1}{3}\left(\abs{g_{5}}^{2}+\abs{g_{6}}^{2}\right)\right.\nonumber\\
	&+\frac{8}{3}\left(\ln\frac{m^2_{\ell_i}}{m^2_{\ell_k}}-3\right)\left(\abs{e^2A_{2R}}^{2}+\abs{e^2A_{2L}}^{2}\right) \nonumber \\
	&- \frac{4}{3}\Re\left(e^2A_{2R}g_{4}^{*}+e^2A_{2L}g_{3}^{*}+e^2A_{2R}g_{6}^{*}+e^2A_{2L}g_{5}^{*}\right)\bigg].
\end{align}
The definitions of effective couplings are the same as the one given in Eq.~\eqref{eq:wc}, with some modification
\begin{align}
	e^2B_{R} =&~ \frac{3}{16\pi^2} \frac{\zeta_a\zeta_b v_{EW}^2}{M_a^2} \lambda^R_{3i}\lambda^{R\ast}_{3j}|\lambda^R_{3k}|^2 b(x_{ta},r_{ba}), \nonumber \\
	e^2B_{L} =& \frac{3}{16\pi^2} \frac{v_{EW}^2}{M_3^2} \lambda_{3i}\lambda^{\ast}_{3j}|\lambda_{3k}|^2b(0,1).
\end{align}

The $Z$-loop corrections, $Z_{L,R}$,  are proportional to $(m_q/m_Zs_Wc_W)^2$. In the case of $Z_L$, it is suppressed by the bottom-quark mass, so it is practically zero. In contrast, $Z_R$ gains a significant enhancement by a factor of 16 from the top mass. This indicates that $Z$ contributions with top exchange can dominate over photon and box diagrams, especially when $\lambda^u_{31}$ or $\lambda^u_{32}$ is very small (no internal chirality enhancement), due to neutrino oscillation constraints. 

However, when these couplings are sufficiently large---a condition typically outside the region allowed by neutrino oscillation data---the corresponding dipole terms start to  become significant. This explains the behaviors of LFV constraints from $\tau\to e\gamma$ and $\tau\to3e$ illustrated in Fig.~\ref{fig:tx2}. Since the $\tau\to 3e$ amplitude has an additional power of electromagnetic coupling compared to $\tau \to e \gamma$,  it is $\tau\to e\gamma$ that provides better constraints on the parameter space for that region of $\lambda^u_{31}$ value. We observe an analogous situation when comparing $\tau\to\mu\gamma$ and $\tau\to3\mu$ decays.

\subsubsection{$\mu-e$ conversion in nuclei}
\label{sec:mue-conv} 
Another bound that may affect the parameter space of the model is the one coming from the conversion of muon to electron in nuclei. This process bears similarity with the $\mu\to e\gamma$ process but with off-shell photon converted into $q\bar{q}$ pair. In addition, there are also $Z$-mediated and  box diagrams, as well as tree-level diagrams. We will be particularly interested in the so-called coherent process, where the final state of the nucleus is the same as the initial one. At the the quark level, the effective Lagrangian is written as~\cite{Kitano:2002mt}
\begin{align}
\mathcal{L}^{eff}_{\mu-e~\text{conv}} =&~ \frac{1}{2v_{EW}^2} \bigg[ em_\mu \bar{e}\sigma^{\mu\nu}A_{2R}P_R \mu F_{\mu\nu} -g_{LS}^q(\bar{e}P_R\mu)(\bar{q}q) - g_{LV}^q (\bar{e}\gamma^\mu P_L \mu)(\bar{q}\gamma^{\mu}q) \bigg]  + (L\leftrightarrow R) + \hc
\end{align}
Note that we do not include pseudoscalar, axial-vector, and tensor quark operators as they do not contribute to the coherent process.  

Considering the coupling textures presented, only $\lambda$ couplings (via $R^{2/3}$ exchange) can induce the $\mu-e$ transition. However, as we have seen before, the corresponding dipole coefficients suffer from the GIM-like cancellation, rendering them negligible. Similarly, the scalar operators cannot be induced because this particular couplings can only connect muon and electron with the same chirality.  

Consequently, only the left-handed vector operator contributes to the process. This operator needs to be matched to the nucleon level, giving $\bra{N}\bar{q}\gamma^\mu q\ket{N}=n^{(q,N)}\bar{N}\gamma^\mu N$, with $n^{(q,N)}$ denoting the number of valence quark $q$ in nucleon $N$. As a result, box and tree-level exchanges can be omitted as they contain only $b$-quark flavor. The effective coupling is given by
\begin{align}
	g_{LV}^q =&~ e^2Q_qA_{1L} -e^2\left[T_3(q)-2Q_qs_W^2\right]Z_L,
	\label{eq:gLV}
\end{align}
where $g_{RV}^q=g_{RS}^q=g_{LS}^q=0$, and the coefficients $A_{1L}$ and $Z_L$ are defined in Eq.~\eqref{eq:wc}. 

From here, the conversion rate is found as~\cite{Kitano:2002mt}
\begin{align}
	\Gamma(\mu-e~\text{conv}) =&~ \frac{m_\mu^5}{v_{EW}^4} \bigg| (2g_{LV}^u +g_{LV}^d)V^p  + (2g_{LV}^d+g_{LV}^u)V^n \bigg|^2,
\end{align}
with $V^{p,n}$ being overlap integrals for the proton and neutron. Their definitions and values for various atomic nuclei are given in Ref.~\cite{Kitano:2002mt}. It is worth noting that the $Z$ contribution is suppressed by $m_b^2/m_Z^2$, so it would induce merely a negligible correction. The dominant contribution is driven by the photon exchange, which is proportional to the electric charge of the valence quarks. Because of that, the term associated with the neutron will be effectively zero. 

Today, the strongest constraint of such a conversion comes from the search in the gold nucleus by SINDRUM II Collaboration~\cite{SINDRUMII:2006dvw}, i.e., $R_{\mu e}\equiv\Gamma(\mu-e~\text{conv})/\Gamma(\text{captured})<7\times 10^{-13}$, with $\Gamma(\text{captured})=13.06\times 10^{-6}\unit{s}^{-1}$ and $V^p=0.0974$. Future experiments, such as COMET~\cite{COMET:2018auw} and Mu2e~\cite{Mu2e:2014fns}, are projected to reach the sensitivities of the order $\mathcal{O}(10^{-18})$ using aluminum target.

\subsection{Constraints from $Z\to \ell^+\ell^-$ decays}
\label{sec:Zdecay}
\label{sec:Zll}
Leptoquark Yukawa interactions, given in Eq.~\eqref{eq:lag-mass}, can induce corrections to the $Zf\bar{f}$ couplings. These interactions are described by the Lagrangian
\begin{align}
    \mathcal{L}_Z = \frac{g}{c_W} \bar{f}\gamma^\mu(g_L^{f}P_L + g_R^{f}P_R)f Z_\mu + \hc,
\end{align}
where $g_{L,R}^{f}\equiv g_{L,R}^{f,SM}+\delta g_{L,R}^f$ represent the effective left- and right-handed couplings.

We are particularly interested in corrections to leptonic couplings $g_{L,R}^{\ell}$, determined from $Z\to \ell^+\ell^-$ decays. These decays have been precisely measured by the LEP experiment, giving the following values~\cite{ALEPH:2005ab}:
\begin{align}
    g_L^{e} =&~ -0.26963\pm0.00030, \quad g_R^{e} = 0.23148\pm0.00029 \nonumber \\
    g_L^{\mu} =&~ -0.2689\pm0.0011, \quad g_R^\mu = 0.2323\pm0.0013 \nonumber \\
    g_L^\tau =&~ -0.26930\pm0.00058, \quad g_R^\tau = 0.23274\pm0.00062.
\end{align}
Corrections to these couplings arise at one-loop level from $Z$-penguin diagrams, in the similar manner as $\ell_i\to3\ell_j$ decays. The difference is here there is a significant correction from the $Z$ momentum, leading to~\cite{Arnan:2019olv}
\begin{align}
    \delta g_L^{\ell} =&~ \frac{|\lambda_{3\ell}|^2x_{Z3}}{16\pi^2}\bigg[g_R^{d}\left(\tfrac{1}{6}+i\pi -\ln x_{Z3} \right) +  \tfrac{1}{6}g_L^{\ell}\bigg], \nonumber \\
    \delta g_R^{\ell} =&~ \frac{3|\lambda^R_{3\ell}|^2\zeta_a}{16\pi^2}\bigg[(g_L^{u^c}-g_R^{u^c})F(x_{ta}) + \tfrac{1}{12}x_{Za}f(x_{ta}) \bigg] + \sum_{i=1,2}\frac{|\lambda^R_{i\ell}|^2\zeta_ax_{Za}}{16\pi^2}\bigg[g_L^{u^c}\left(\tfrac{1}{6} + i\pi - \ln x_{Za} \right) +  \tfrac{1}{6}g_R^{\ell}\bigg],
\end{align}
with $x_{Z3}=m_Z^2/M_3^2$ and
\begin{align}
    f(x) =&~ g_{L}^{u^c} \frac{(1-x)(-8+7x-5x^2)-2(2+x^3)\ln x}{(1-x)^4} + g_{R}^{u^c} \frac{(1-x)(2+5x-x^2)+6x\ln x}{(1-x)^4} \nonumber \\
    &~ + g_{R}^{\ell} \frac{(1-x)(2-7x+11x^2)+6x^3\ln x}{3(1-x)^4}.
\end{align} 
Again, contributions from $\lambda^u_{ij}$ couplings are ignored, and all fermion masses except the top mass are set to zero.

\subsection{Electric dipole moment of electron}
\label{sec:edm}
Another important observable for our discussion is the electric dipole moment of electron (electron EDM) $d_e$, defined through the Lagrangian term
\begin{align}
    \mathcal{L} = -\frac{d_e}{2}\left(\bar{e}\sigma^{\mu\nu}i\gamma_5e\right)F_{\mu\nu}.
    \label{eq:lag-edm}
\end{align}
Within this model, such a quantity can only be induced at one-loop level via internal chirality flip and is given by
\begin{align}
    d_e = \frac{3e}{16\pi^2} \frac{\zeta_a m_{u_k}}{M^2_a}\Im(\lambda^R_{k1}\lambda^{u\ast}_{k1})\kappa'(x_{ka}).
\end{align}
The current limit on this quantity is $|d_e|<4.1\times 10^{-30}e$~cm~\cite{ParticleDataGroup:2024cfk}. For a specific $M_1=1500~\text{GeV}$, $\Delta M=50~\text{GeV}$, and maximal LQ mixing, the experimental bound on $d_e$ translates to 
\begin{align}
    |\Im(\lambda^R_{21}\lambda^{u\ast}_{21})| <&~5.0\times 10^{-9}~~ \text{for~TX~1}, \nonumber \\
    |\Im(\lambda^R_{31}\lambda^{u\ast}_{31})| <&~8.4\times 10^{-11}~~ \text{for~TX~2}. 
\end{align}
These stringent limits suggest either extremely tiny values for $\lambda^R_{21},\lambda^R_{31}$ or highly aligned phases such that $\arg(\lambda^R_{k1})=\arg(\lambda^u_{k1})+n\pi$, where $n$ is an arbitrary integer. Similar observations are also seen in the case of the muon EDM. 

\subsection{High-$p_T$ dilepton tails}
\label{sec:dilepton}
Some couplings within this model can also contribute to dilepton production $pp\to \ell^+\ell^-$ at the LHC, i.e., via the tree-level LQ exchanges. This finds its relevance when the corresponding quarks are of first or second generation. Since the LQ masses we consider are well above the electroweak scale, we can adopt the effective theory approach. We are particularly interested with the following flavor-conserving dim-6 operators involving right-handed up-type quark
\begin{align}
	\mathcal{L}^{eff} = -\frac{2}{v_{EW}^2}C^{RR}_{ij} (\bar{q}_i\gamma^\mu P_R q_i)(\bar{\ell}_j\gamma_\mu P_R\ell_j),
\end{align}
with
\begin{align}
	C^{RR}_{ij}=~ \frac{v_{EW}^2}{4}\left(\frac{c_\theta^2}{M_1^2} + \frac{s_\theta^2}{M_2^2}\right)|\lambda^R_{ij}|^2.
\end{align}
Other operators with left-handed quarks or leptons, i.e., $C^{LL}_{ij}$ or $C^{RL}_{ij}$, are not considered. This is  because these operators involve $\lambda^u$ couplings, which are constrained to be small by neutrino oscillation data. 

Two particular coefficients that are relevant for our discussion are $C^{RR}_{21}$ in TX~1 and $C^{RR}_{22}$ in TX~2. The magnitudes of these coefficients have been tabulated in Ref.~\cite{Greljo:2017vvb} based on high-$p_T$ dilepton search by ATLAS Collaboration. Their conservative bounds are found to be $|C^{RR}_{21}|<0.0127~(0.00417)$ and $|C^{RR}_{22}|<0.0162~(0.00632)$ for $36.1~\text{fb}^{-1}~(3000~\text{fb}^{-1})$ luminosity.

\subsection{Flavor constraints in the quark sector}
\label{sec:had}
So far, we have only discussed the effects on LQ interactions in the lepton sector. In principle, these interactions may induce flavor-changing processes in the quark sector. However, in general, those effects are negligible. This is primarily due to the suppression of $\lambda^u$ elements by neutrino data, which are related to down-type couplings via $\lambda^d=V^T\lambda^u$, and the TeV-scale of LQ masses. In addition, texture choices of Yukawa coupling matrices, $\lambda$ and $\lambda^R$, allow us to explicitly avoid certain flavor-changing processes.

To illustrate this, let us examine the $b\to s\nu\bar{\nu}$ transition. In the SM, such a $\Delta B=1$ (not to be confused with baryon number $B$ mentioned in earlier section) process is expected to be suppressed, as its leading contributions only arise at one-loop level via $Z$-penguin and box diagrams.  Within the present LQ framework, it can appear at tree level driven by $\lambda^d$ couplings. (Another tree-level contribution induced by $\lambda$ does not occur because, in the present discussion, this coupling matrix only couples to the $b$-quark.) The corresponding effective Hamiltonian can be expressed as
\begin{align}
\mathcal{H}_{\rm eff} = -\frac{4G_F}{\sqrt{2}}V_{tb}V_{ts}^\ast C^\nu_L \mathcal{O}_L^\nu, 
\end{align}  
where
\begin{align}
	\mathcal{O}_{L}^\nu = \frac{\alpha_{em}}{4\pi}\left(\bar{s}\gamma^\mu P_{L} b\right)\left(\bar{\nu}_i\gamma_\mu(1-\gamma_5)\nu_j \right).
\end{align}
and $C^\nu_L\equiv C^{\nu,\text{SM}}_L+C^{\nu,\text{new}}_L$. The SM Wilson coefficient is given by $C^{\nu,\text{SM}}_L=-X(m_t^2/m_W^2)/s_W^2$, with the function $X(x)$ defined in~\cite{Misiak:1999yg,*Buchalla:1998ba}, whereas the new contribution is found as
\begin{align}
	C^{\nu,\text{new}}_L = -\frac{\sqrt{2}\pi \lambda^{d}_{3i}\lambda^{d\ast}_{2j}}{4G_FM_{LQ}^2V_{tb}V_{ts}^\ast\alpha_{em}}.
\end{align}
In deriving the above relation, we have assumed a maximal LQ mixing angle and $M_1\simeq M_2=M_{LQ}$.
Numerically, $C^{\nu,\text{SM}}_L=-6.38\pm0.06$~\cite{Altmannshofer:2009ma}. Using the coupling values from Fig.~\ref{fig:g-2} with $M_{LQ}=1.5$ TeV, it follows that $|C^{\nu,\text{new}}_L/C^{\nu,\text{SM}}_L|\sim \mathcal{O}(10^{-8})$ for both TX~1 and TX~2, indicating that the $b\to s\nu\bar{\nu}$ transition receives no significant correction from the LQ sector. Consequently, the rates for related processes, such as $B\to K\nu\bar{\nu}$, remain consistent with SM predictions.  The suppression of NP effects in kaon decays can also be explained in an analogous way. 

We now turn to the $b\to s\ell^+\ell^-$ transition, whose effective Hamiltonian is stated as
\begin{align}
	\mathcal{H}_{\rm eff} = -\frac{4G_F}{\sqrt{2}}V_{tb}V_{ts}^\ast\left[ C_9^{\ell\ell}\mathcal{O}_9^{\ell\ell} + C_{10}^{\ell\ell'}\mathcal{O}_{10}^{\ell\ell} \right],
\end{align}
with $\mathcal{O}_{9(10)}^{\ell\ell}=(\alpha_{em}/4\pi)[\bar{s}\gamma^\mu P_Lb][\bar{\ell}\gamma_\mu(\gamma_5)\ell]$ and $C_{9(10)}^{\ell\ell}=C_{9(10)}^{\ell\ell,\text{SM}}+C_{9(10)}^{\ell\ell,\text{new}}$. For the SM, we have $C_{9}^{\ell\ell,\text{SM}}=-C_{10}^{\ell\ell,\text{SM}}=Y(m_t^2/m_W^2)/s_W^2$, with the expression of $Y(x)$ given in \cite{Misiak:1999yg,*Buchalla:1998ba}. The LQ contributions arise through box diagrams, given by
\begin{align}
	C_9^{\ell\ell,\text{new}} = C_{10}^{\ell\ell,\text{new}} =-\frac{\sqrt{2} (\lambda^{d}\lambda^{d\dagger})_{32}(\lambda^{R\dagger}\lambda^R)_{\ell\ell}}{128\pi G_FM_{LQ}^2V_{tb}V_{ts}^\ast\alpha_{em}}.
\end{align}
Although $\lambda^R\sim\mathcal{O}(1)$, these new contributions are still orders of magnitude smaller than the SM values. Thus, no significant deviations from the SM predictions are expected for $B\to K\ell^+\ell^-$ and $B_s\to\ell^+\ell^-$ decays. Furthermore, decays into different light lepton flavors, e.g., $B\to Ke^\pm\mu^\mp$, can be trivially satisfied, thanks to the decoupling nature of the electron and muon sectors. Also, the new effects on $\Delta B=2$ and $\Delta S=2$ processes, such as $B_{d,s}-\bar{B}_{d,s}$ and $K-\bar{K}$ mixings, are expected to be more suppressed due to having more power of $\lambda^d_{ij}$ in their diagrams. 

The above analysis can be extended to cases involving charm quarks, relevant for $B\to D\tau\nu$ or $D\to \ell\nu$ decays. We start with $b\to c\tau\nu$ transition, whose effective Hamiltonian is given by
\begin{eqnarray}
	\mathcal{H}_{\text{eff}}&=&\frac{4G_{F}}{\sqrt{2}}V_{cb}\left[(1+C_{L}^{V})(\bar{c}\gamma^{\mu}P_{L}b)(\bar{\tau}\gamma_{\mu}P_{L}\nu) +C_{L}^{S}(\bar{c}P_{L}b)(\bar{\tau}P_{L}\nu)+C_{L}^{T}(\bar{c}\sigma^{\mu\nu}P_{L}b)(\bar{\tau}\sigma_{\mu\nu}P_{L}\nu)\right],
\end{eqnarray}
with Wilson coefficients
\begin{eqnarray}
	C_{L}^{V}&=&\frac{\lambda^{d}_{3i}\lambda^{u\ast}_{23}\sqrt{2}}{8G_{F}V_{cb}M_{\text{LQ}}^{2}}, \quad 
	C_{L}^{S}= -4C_L^T=-\frac{\lambda^{d}_{3i}\lambda^{R*}_{23}\sqrt{2}}{8G_{F}V_{cb}M_{\text{LQ}}^{2}}.
\end{eqnarray}
Again, no deviations from SM values are expected to occur: $|C_L^V|=0$, thanks to $\lambda^u_{23}=0$ in these specific textures, while $\lambda^R_{23}\sim 1$, although allowed by neutrino oscillation data, cannot compensate for the suppression of $\lambda^d_{3i}$, resulting in $|C_L^S|\ll 1$. The same behavior is also observed in decays involving light leptons $b\to c\ell\nu$ ($\ell=e,\mu$) and $D\to\ell\nu$. The former, in particular, leads to $R_{D^{(\ast)}}\equiv \text{BR}(B\to D^{(\ast)}\tau\nu)/\text{BR}(B\to D^{(\ast)}\ell\nu)$ values that remain consistent with the SM.

%%%%%%%%%%%%%%%%%%%%%%%%%%%%%%%%%%%%%%%%%%%%%%%
%%%%%%%%%%%%%%%%%%%%%%%%%%%%%%%%%%%%%%%%%%%%%%%
\section{Conclusions}\label{sec:concl}
In this paper, we analyzed a radiative neutrino mass model featuring two scalar leptoquarks $S(3,1,-1/3)$ and $R(3,2,1/6)$ in the context of recent lepton $g-2$ observations. By employing flavor textures that decouple the electron and muon sectors, we ensured that the anomalous magnetic dipole moments are induced by different up-type quarks, thereby suppressing dangerous $\mu\to e\gamma$ decay. In the minimal scenario, consistency with neutrino oscillation data requires the simultaneous presence of one- and two-loop neutrino mass contributions. 

The model predicts a massless lightest neutrino, accommodates both normal and inverted orderings, and establishes a direct link between neutrino oscillation data and the lepton $g-2$ values. We find that while neutrino data prevent a large new correction to the muon $g-2$, the electron $g-2$ discrepancy (as implied by the Rb experiment) can be resolved within $2\sigma$, achievable only for the inverted neutrino mass ordering. Furthermore, our analysis indicates that the model remains perturbative and technically natural up to 100~TeV. Finally, the predicted rates for LFV processes, such as $\tau\to e\gamma$ and $\tau\to3e$, lie near current experimental limits.  Future results from neutrino oscillations, high-luminosity colliders, and LFV searches will provide critical tests to confirm or falsify this scenario.

%%%%%%%%%%%%%%%%%%%%%%%%%%%%%%%%%%%%%%%%%%%%%%%%%%
\begin{acknowledgments}
The work of B. D. was supported in part by the National Research and Innovation Agency (BRIN) through Research Assistantship Program (Contract No. 3/II/HK/2022).
\end{acknowledgments}

%%%%%%%%%%%%%%%%%%%%%%%%%%%%%%%%%%%%%%%%%%%%%%%
\bibliographystyle{apsrev4-1}
\bibliography{references}
%%%%%%%%%%%%%%%%%%%%%%%%%%%
\end{document}